\documentclass[a4paper,fleqn,usenatbib]{mnras}

\usepackage{newtxtext,newtxmath}
\usepackage{tabularx}
\usepackage{multirow}
\usepackage{makecell}
\usepackage{hhline}
\usepackage{currvita}

\usepackage[T1]{fontenc}
\usepackage{ae,aecompl}

\usepackage{graphicx}	
\usepackage{amsmath}	
\usepackage{amssymb}	
\usepackage{subfigure}
\usepackage[left]{lineno}
\usepackage{eso-pic}
\usepackage[multiple]{footmisc}

\newcommand{\NOOP}[1]{} 
\newcommand{\COMMENT}[1]{\NOOP{#1}}
\newcommand{\NEW}[1]{#1}

\title[Star-galaxy classification in DESY1 data]{Star-galaxy classification in the Dark Energy Survey Y1 dataset}

\author[I.~Sevilla-Noarbe et al.]{
\parbox{\textwidth}{I.~Sevilla-Noarbe$^{1}$\thanks{E-mail: ignacio.sevilla@ciemat.es}, B.~Hoyle$^{2,3}$, M.J.~March\~a$^{4}$, M.T.~Soumagnac$^{5}$,
K.~Bechtol$^{6}$, A.~Drlica-Wagner$^{7}$, F.~Abdalla$^{4,8}$, J.~Aleksi\'c$^{9}$, C.~Avestruz$^{10}$, E.~Balbinot$^{11}$, M.~Banerji$^{12,13}$, E.~Bertin$^{14,15}$, C.~Bonnett$^{9}$, R.~Brunner$^{16}$, M.~Carrasco-Kind$^{17}$, A.~Choi$^{18}$, T.~Giannantonio$^{12,13,2}$, E.~Kim$^{17}$, O.~Lahav$^{4}$, B.~Moraes$^{4}$, B.~Nord$^{7}$, A.J.~Ross$^{18}$, E.S.~Rykoff$^{19,20}$, B.~Santiago$^{21,22}$, E.~Sheldon$^{23}$, K.~Wei$^{10,24}$, W.~Wester$^{7}$, B.~Yanny$^{7}$, T.~Abbott$^{25}$, S.~Allam$^{7}$, D.~Brooks$^{4}$, A.~Carnero-Rosell$^{22,26}$, J.~Carretero$^{9}$, C.~Cunha$^{19}$, L.~da Costa$^{22,26}$, C.~Davis$^{19}$, J.~de Vicente$^{1}$, S.~Desai$^{27}$, P.~Doel$^{4}$, E.~Fernandez$^{9}$, B.~Flaugher$^{7}$, J.~Frieman$^{7,10}$, J.~Garcia-Bellido$^{28}$, E.~Gaztanaga$^{29,30}$, D.~Gruen$^{19,20}$, R.~Gruendl$^{16,17}$, J.~Gschwend$^{22,26}$, G.~Gutierrez$^{7}$, D.L.~Hollowood$^{31}$, K.~Honscheid$^{18,32}$, D.~James$^{33}$, T.~Jeltema$^{31}$, D.~Kirk$^{4}$, E.~Krause$^{34,35}$, K.~Kuehn$^{36}$, T.~S.~Li$^{7,10}$, M.~Lima$^{37,22}$, M.~A.~G.~Maia$^{22,26}$, M.~March$^{38}$, 
R.~G.~McMahon$^{4,8}$, F.~Menanteau$^{16,17}$, R.~Miquel$^{39,9}$, R.~L.~C.~Ogando$^{22,26}$, A.~A.~Plazas$^{35}$, E.~Sanchez$^{1}$, V.~Scarpine$^{7}$, R.~Schindler$^{20}$, M.~Schubnell$^{40}$, M.~Smith$^{41}$, R.~C.~Smith$^{25}$, M.~Soares-Santos$^{42}$, F.~Sobreira$^{43,22}$, E.~Suchyta$^{44}$, 
M.~E.~C.~Swanson$^{17}$, G.~Tarle$^{40}$, D.~Thomas$^{45}$, D.~L.~Tucker$^{7}$, A.~R.~Walker$^{25}$
\begin{center} (The DES Collaboration) \end{center}
}
\vspace{0.4cm}
\\
\parbox{\textwidth}{\rm Author affiliations are listed at the end of this paper.}
\\
\\
}

\date{Accepted XXX. Received YYY; in original form ZZZ}

\pubyear{2018}

\hypersetup{draft}

\begin{document}

\COMMENT{\AddToShipoutPictureBG*{%
  \AtPageUpperLeft{%
    \hspace{0.75\paperwidth}%
    \raisebox{-3.5\baselineskip}{%
      \makebox[0pt][l]{\textnormal{DES-2017-0227}}
}}}%
\AddToShipoutPictureBG*{%
  \AtPageUpperLeft{%
    \hspace{0.75\paperwidth}%
    \raisebox{-4.5\baselineskip}{%
      \makebox[0pt][l]{\textnormal{Fermilab-PUB-18-112-AE-PPD}}
}}}%
}

\label{firstpage}
\pagerange{\pageref{firstpage}--\pageref{lastpage}}
\maketitle

\begin{abstract}
We perform a comparison of different approaches to star-galaxy classification using the broad-band photometric data from Year 1 of the Dark Energy Survey. \COMMENT{We explore and compare different techniques to address this problem: morphological versus using additional flux information, differences between machine-learning, cut based and template fitting including Hierarchical Bayesian methods and catalogue-based versus image-based analyses. We use} This is done by performing a wide range of tests with and without external `truth' information, which can be ported to other similar datasets. We make a broad evaluation of the performance of the classifiers in two science cases with DES data that are most affected by this systematic effect: large-scale structure and Milky Way studies. In general, even though the default morphological classifiers used for DES Y1 cosmology studies are sufficient to maintain a low level of systematic contamination from stellar mis-classification, contamination can be reduced to the O(1\%) level by using multi-epoch and infrared information from external datasets. For Milky Way studies the stellar sample can be augmented by $\sim20\%$ for a given flux limit. Reference catalogues used in this work are available at \NEW{\url{http://des.ncsa.illinois.edu/releases/y1a1}}. 
\end{abstract}




\section{Introduction}

Accurate classification of astrophysical sources is essential for interpreting photometric surveys. Specifically, separating foreground stars from background galaxies is important for many astronomical research topics, from Galactic science to cosmology. Conventional morphological classification techniques separate point sources (mostly stars) from resolved sources (galaxies) using selections in magnitude-radius space or similar variables \citep{macg76, kron80, heyd89, yee91}. For bright sources, morphology has proven to be a sufficient metric for classification. In this regime, for weak lensing applications, a very pure, but also abundant, star sample is vital for deriving the correct point spread function in the images which is used to later infer cosmic shear \citep{soumagnac,jarvis,zuntz}. At fainter magnitudes unresolved galaxies will begin to contaminate catalogues of point-like sources and noisy measurements of stars will contaminate the galaxy sample. Blended sources become an issue as well, because distant and/or faint sources start to merge into single detected objects with spurious shapes. Mis-classification of stars and galaxies at faint magnitudes can introduce spurious correlations in galaxy surveys \citep{ross} and will hamper the study of stellar distributions \citep{y2satellites}. 


The advent of CCD detectors provided larger, more reliable data sets which became an obvious target for machine learning classification algorithms \citep[e.g.,][]{odewahn,sextractor,bdts,mlexplore,kimdcnn}. In addition, many large, multi-band imaging surveys \NEW{use morphology \citep[such as for SDSS,][]{stoughton} and/or} have incorporated colour information into their classifiers (see \citealt{ball} for SDSS \NEW{as well}, \citealt{hildebrandt} for CFHTLS or \citealt{panstarrs} for Pan-STARRS). Adopting a Bayesian approach to incorporate fits to stellar and galaxy templates has been shown to be a promising avenue \citep{fadely}, as well as the use of infrared data to complement the optical band observations \citep{malek,kovacs,banerji}.


In this paper we test different strategies for classifying objects as point-like or extended sources in the Dark Energy Survey (DES) Year 1 data (Y1). We subsequently analyze the impact in two broad science cases, and possible developments to improve object classification in future analyses of this data. Throughout this paper, `extended' will be used as a synonym for `galaxy' whereas `point-like' includes both stars and quasi-stellar objects (QSOs) on first approximation and we will collectively call them `stars' in this work. For the case studies considered here and the general catalogue, the contamination of QSOs in the large-scale stellar and galactic catalogues is not deemed important. However, a good star-QSO separation is needed for quasar science, as studied in detail in \citet{tiess} for DES data.
 



After a description of the dataset in Section \ref{sec:dataset} and the classifiers we are considering here in Section \ref{sec:classifiers}, we compare the classifiers in calibration fields (Section \ref{sec:calib}) and then analyze the response in the complete Y1 dataset for a few selected ones (Section \ref{sec:y1valid}). Then we study the impact on large-scale structure and Milky Way studies (Section \ref{sec:discussion}). Finally, Section \ref{sec:conclusions} presents the conclusions and discusses possible additional developments.

\section{Dark Energy Survey datasets}
\label{sec:dataset}
The DES consists of a 5000 square-degree "wide" survey using the \textit{grizY} photometric bands to AB 10$\sigma$ magnitude limits of (24.6, 24.4, 23.7, 22.7, 21.5) respectively for 2 arc-second apertures, together with a ${\sim}27$ square degree supernovae survey observed in the \textit{griz} bands  with an approximately weekly cadence.  In February 2018, the project completed the original five planned observing seasons (Years 1 through 5, Y1-Y5). Additional science-quality data was collected during an earlier Science Verification (SV) season. The core goal of DES is a multi-probe study of dark energy at different cosmological epochs using the same DECam instrument \citep{decam} and DES Data Management (DESDM) pipeline \citep{desdm}, as showcased with its first results in \citet{desy1cosmo}. However, the richness of this dataset allows astronomers and cosmologists to go beyond this initial objective \citep{morethande}. 

For this study, we use the subset of highest quality data from DES SV\footnote{\url{https://des.ncsa.illinois.edu/releases/sva1}} and Y1\NEW{\footnote{\url{https://des.ncsa.illinois.edu/releases/y1a1}}}\citep{y1gold} comprising the "Gold" catalogue. 
We note the following features that are relevant for the present study:

\begin{itemize}
\item The object catalogues are obtained applying \texttt{SExtractor} \citep{sextractor} to coadded images with typically 2 to 4 overlapping exposures in each band in the case of Y1 or $\sim 10$ for SV data, with object detection performed on a combined \textit{riz} image. 
\item \texttt{SExtractor} magnitudes have been calibrated through a global calibration module \citep{tucker} and subsequently adjusted through a fit to the stellar locus \citep{slr} anchored to the \textit{i} band\footnote{This calibration approach was eventually superseded in Y3 data products with the Forward Global Photometric Calibration approach described in \cite{fgcm}}. This procedure also corrects for Galactic extinction. In general, \texttt{MAG\_AUTO} is used for photometry (for binning purposes and as inputs for the template based method described below), as it behaves more robustly for these coadded catalogues. \texttt{MAG\_MODEL}, \texttt{MAG\_DETMODEL}\footnote{In this case the exponential model used in \texttt{SExtractor} is fitted on the detection image and scaled in the measurement images of each band.} and \texttt{MAG\_PSF} are used as inputs for the machine learning methods as well. \NEW{Shape measurements in this code include \texttt{FLUX\_RADIUS, CLASS\_STAR and SPREAD\_MODEL}, some of which will be specifically studied here.}
\item In addition, a multi-object, multi-epoch fitting pipeline (\texttt{MOF}) has been run on the single-epoch image counterparts for each coadd catalogue detection to obtain more precise photometric measurements for the objects. \NEW{It simultaneously fits a Gaussian mixture model to} the individual images, also modelling light from nearby neighbours for each object (more details in \citealt{y1gold}). \NEW{The main flux measurements used for the methods described here are the fluxes using this composite Gaussian mixture model (\texttt{CM\_MAG}) and the PSF magnitudes derived from the same \texttt{MOF} pipeline (\texttt{PSF\_MAG}). \texttt{CM\_T} is a size estimator from the code before PSF convolution, which will be studied here in detail.}
\item All objects are required to be in areas for which there is at least one exposure in each of the \textit{griz} bands. 
\end{itemize}

We define two distinct regions in which we will perform our tests:

\begin{enumerate}
\item A \textbf{calibration field}: defined by those areas that overlap external datasets that we can use to train, validate and test our methods. These are the supernova (SN) fields from the DES SN survey, which overlap specific spectroscopic surveys and miscellaneous Hubble Space Telescope (HST) datasets; and the area of the survey overlapping the Sloan Digital Sky Survey \citep[SDSS;][]{sdss} Stripe 82 region \citep{sloansn}. In addition, the COSMOS field\footnote{\url{http://cosmos.astro.caltech.edu/}} has been imaged with DECam, providing a very useful dataset given the richness of multi-band imaging and spectroscopy available. Table \ref{tab:external_datasets} summarises the numbers of objects matched to various external datasets (details in Section \ref{sec:train_test_fields}).
Some of these fields have a large number of DES exposures, due to their application for SN searches, so special coadds were made from $\sim4$ exposures in each band in order to resemble the Y1 depth. The selection of these exposures was made so that their coaddition would provide similar characteristics in terms of sky brightness and seeing as the wide survey coadds \citep{y1gold,neilsen}. This procedure is not needed in forthcoming releases as the wide survey extends to cover the supernova regions. 
\item An \textbf{application field}: the remaining area of the DES footprint for which suitable external datasets for training are not presently available. This includes the so-called `SPT' region due to the overlap with the South Pole Telescope\footnote{\url{https://pole.uchicago.edu/}} \citep{spt} observations, in which we can make some quality assessment as well, though limited by the lack of external references. 
\end{enumerate}

\begin{table*}
	\centering
	\caption{External datasets used in this work. \NEW{SDSS-stripe 82 data shows two numbers according to simultaneous 2MASS and WISE matches, and VHS matches. More details are provided in Appendix \ref{sec:external_datasets}}.}
	\label{tab:external_datasets}
	\begin{tabular}{ccccc} 
		\hline
		Catalogue & Type & Usage in this work & Nb. matched objects & Reference    \\ 
		\hline
        ACS-COSMOS & Space optical imaging & Truth table & 116017 & \citet{leauthaud} \\
        Hubble-SC & Space optical imaging & Truth table & 12927 & \citet{hsc} \\
        SDSS-stripe 82 & Ground optical spectroscopy & Truth table & 18984/46700 & \citet{sdssdr13} \\
        VVDS & Ground optical spectroscopy & Truth table & 4442 & \citet{vvds} \\
        WISE & Space NIR imaging & Complementary data & 18984 & \citet{wise} \\
        2MASS & Ground NIR imaging & Complementary data & 18984 & \citet{2mass} \\  
        VHS & Ground NIR imaging & Complementary data & 46700 & \citet{vhs} \\
		\hline
	\end{tabular}
\end{table*}

\section{Description of the object classifiers}
\label{sec:classifiers}

\begin{table*}
	\centering
	\caption{Summary of classification methods. \NEW{Type of data denotes whether measurements or direct pixel data are used, and in the first case if it is based on morphological and/or flux measurements. The specific algorithmical approach is named on the third column.}}
	\label{tab:classifiers}
	\begin{tabular}{ccc} 
		\hline
		Name & Type of data used & Algorithm \\
		\hline
		CLASS\_STAR & Isophotal level measurements, morphological &  Neural Network \\
		SPREAD\_MODEL & Pixel-level & Normalised Linear Discriminant \\
        CM\_T & Measurements on fitted shape, morphological & Second moments of Gaussian mixture fit (object)\\
        MCAL\_RATIO & Measurements on fitted shape, morphological & 2nd moments of Gaussian mixture fit (noisified object and PSF) \\
        ADA\_PROB & \NEW{\makecell{Most discriminating features \\ from a combination of simple functions \\ used over all catalogue columns.}} & Boosted Decision Trees \\
        GALSIFT\_PROB & All catalogue columns (PCA) & Random Forests \\
        SVM & \makecell{\NEW{\texttt{MAG\_AUTO},\texttt{FLUX\_RADIUS},\texttt{SPREAD\_MODEL}}, \\ \NEW{flux and morphology}} & Support Vector Machine \\
        CONCENTRATION & Catalogue information, morphological & Direct subtraction of magnitudes measured with model and PSF\\
        W1-J, J-K & Catalogue information, fluxes & Colour cut \\
        HB\_PROB & Catalogue information, fluxes & Template fitting of \NEW{spectral energy distributions} \\
		\hline
	\end{tabular}
\end{table*}

Table \ref{tab:classifiers} summarises the methods explored in this paper to perform object classification. These include a variety of algorithms using machine learning methods (training on morphological and/or colour information), pixel-level flux measurements and template-fitting. For the sake of clarity and conciseness, not all algorithms are subjected to every test in this paper, but usually a selection is made in each case. Additional details and references are given below:

\subsection{CLASS\_STAR}
This is the standard \texttt{SExtractor} star-galaxy classifier, providing a neural network real number output (a `stellarity' index from 0 to 1) based on the training on a large simulation of galaxy and star images on CCDs.
\paragraph*{\textit{Input data:}} For every object, eight isophotal areas above the background are measured, plus the value of the intensity at the peak pixel in the object and the value of the FWHM for the image.
\paragraph*{\textit{Method:}} It uses a backpropagation model \citep{werbos} for learning, based on simulations that include a wide range of PSF profiles and sizes, though they are optimised to work best on intermediate magnitude ranges (in the DES magnitude scale) of $V\sim 18-22$ due to the types of galaxies simulated and relative star-galaxy mixture.

\subsection{SPREAD\_MODEL} 
This quantity is a linear discriminant-based algorithm available with the \texttt{SExtractor} package. The \texttt{SPREAD\_MODEL} estimator was originally developed as a star-galaxy classifier for the DESDM pipeline, and has also been used in other surveys \citep[e.g.,][]{desai,bouy}. 
\paragraph*{\textit{Input data:}} the image data at pixel level is used for each detected object in \texttt{SExtractor}.
\paragraph*{\textit{Method:}} \texttt{SPREAD\_MODEL} indicates which of the best fitting local PSF model $\vec{\phi}$ (representing a point source) or a slightly more extended model $\vec{G}$ (representing a galaxy) better matches the image data. $\vec{G}$ is obtained by convolving the local PSF model with a circular exponential model with scale length = 1/16 FWHM (Full-Width at Half-Maximum). \texttt{SPREAD\_MODEL} is normalised to allow comparing sources with different PSFs throughout the field:

\begin{equation}
{\tt SPREAD\_MODEL} = \frac{\vec{G}^T {\bf W}\,\vec{p}}{\vec{\phi}^T {\bf W}\,\vec{p}}
        - \frac{\vec{G}^T {\bf W}\,\vec{\phi}}{\vec{\phi}^T {\bf W}\,\vec{\phi}},
\end{equation}

\noindent where $\vec{p}$ is the image vector centered on the source\footnote{This definition of {\tt SPREAD\_MODEL} differs from the one given in previous papers \citep{desai,bouy}, which was incorrect. In practice both estimators give very similar results.}. ${\bf W}$ is a weight matrix constant along the diagonal except for bad pixels where the weight is 0. By construction, \texttt{SPREAD\_MODEL} is close to zero for point sources, positive for extended sources (galaxies), and negative for detections smaller than the PSF, such as cosmic ray hits. The RMS error on \texttt{SPREAD\_MODEL} is estimated by propagating the uncertainties on individual pixel values:
\begin{eqnarray}
{\tt SPREADERR\_MODEL} & = & \frac{1}{(\vec{\phi}^T {\bf W}\,\vec{p})^2} \left(\vec{G}^T {\bf V}\,\vec{G}\,(\vec{\phi}^T {\bf W}\,\vec{p})^2\right.\nonumber \\
  & & + \vec{\phi}^T {\bf V}\,\vec{\phi}\,(\vec{G}^T {\bf W}\,\vec{p})^2\nonumber \\
  & & \left. - 2 \vec{G}^T {\bf V}\,\vec{\phi}\,(\vec{G}^T {\bf W}\,\vec{p}\, \vec{\phi}^T {\bf W}\,\vec{p}) \right)^{1/2}
\end{eqnarray}
where ${\bf V}$ is the noise covariance matrix, which is assumed to be diagonal.

An example of a classifier derived from {\tt SPREAD\_MODEL} is the default classification scheme (\texttt{MODEST\_CLASS}) used in the Y1 Gold catalogue, which includes the following criteria:
\begin{equation}
\begin{split}
galaxies \iff \\ 
&\texttt{SPREAD\_MODEL\_I} + \\
&(5/3) \times \texttt{SPREADERR\_MODEL\_I} > 0.005 \\
&\texttt{AND NOT} \\
&(|\texttt{WAVG\_SPREAD\_MODEL\_I}| < 0.002 \\
&\texttt{AND}\\
&\texttt{MAG\_AUTO\_I} < 21.5)
\end{split}
\label{eq:modgal}
\end{equation}
\begin{equation}
\begin{split}
stars \iff \\ 
&|\texttt{SPREAD\_MODEL\_I} + \\
&(5/3) \times \texttt{SPREADERR\_MODEL\_I}| < 0.002 \\
\end{split}
\label{eq:modsta}
\end{equation}
where \texttt{WAVG\_SPREAD\_MODEL} has been computed from a weighted average of the \texttt{SPREAD\_MODEL} values of single-epoch shapes corresponding to that coadd object. These provide a better separation \citep{dr1} with respect to the standard \texttt{SPREAD\_MODEL} on coadd images, albeit with a limited depth reach, as not all coadd objects have single epoch detections from which a weighted averaged can be computed (a faint object could be detected \textit{only} in the coadded image and not in the individual epochs contributing to the image). The weights come from the weight map of the Data Management processing outputs and the band chosen is the \textit{i} band where the images have a higher signal to noise, and have also demonstrated best performance in detailed simulations. Objects which do not fall into the categories expressed by Equations \ref{eq:modgal} and \ref{eq:modsta} are grouped into either a `fringe' category between both or an `artifact' category (approximately 5\% of the catalogue considered here).


\subsection{CM\_T} 
\texttt{CM\_T} is an intrinsic size estimator for the object from the image fitting provided by the \texttt{MOF} pipeline. 
\paragraph*{\textit{Input data:}} The fitted Gaussian mixture model using the shapes across the images composing the coadd detection.
\paragraph*{\textit{Method:}}
The \texttt{MOF} code estimates the shapes and fluxes of objects detected in the coadd catalogues, using a mixture of Gaussians\footnote{\url{https://github.com/esheldon/ngmix}}\footnote{\url{https://github.com/esheldon/ngmixer}} to simulate the PSF light profile and then convolve them with assumed bulge and disk models (fitted independently for each object, finding the best linear combination) likewise approximated using Gaussian mixtures \citep{hogg}. This is done by fitting across several images of the same object in multiple epochs and bands and then subtracting the flux of neighbours accurately. Concretely, \texttt{CM\_T} is defined as:
\begin{equation}
{\tt CM\_T} = \langle x^2 \rangle + \langle y^2 \rangle
\label{eq:tsize}
\end{equation}
where $x$ and $y$ denote the distance from the object's centre \NEW{determined by the model fit. The value $\langle x^2 \rangle + \langle y^2 \rangle$ can be obtained analytically from the individual component Gaussians.} The PSF is convolved with the fitted model to obtain these pre-PSF values. An associated uncertainty is computed as well, and our best performing classifier, as tested\footnote{Technically, a different, \textit{validation} set would be required to tune this classifier in terms of the quantity multiplying {\tt CM\_T\_ERR}, to avoid bias a towards a specific value, though in practice the differences are small between different choices.} in the COSMOS field, is based on the quantity ${\tt CM\_T} + 2\times{\tt CM\_T\_ERR}$. Typical values are in the range between -0.5 and 0.5.

\subsection{MCAL\_RATIO}
This measurement is derived from the size estimates obtained by the \textit{metacalibration} technique, developed for shear measurement in weak lensing studies \citep{metacal}, \NEW{in which the single epoch objects are artificially sheared to quantify the response of such an effect in the image.}
\paragraph*{\textit{Input data:}} The object size and PSF model size obtained using this technique.  
\paragraph*{\textit{Method:}} This approach uses the same \texttt{ngmix} code as \texttt{MOF} above. However this measurement is much noisier as the metacalibration technique \citep{huffmandelbaum} adds extra noise as part of the correlated noise correction.  This is part of the procedure to correct for selection effects in shear inference, as detailed in \citet{metacal}. \NEW{The discriminating quantity used is}: 

\begin{equation}
\texttt{MCAL\_RATIO} = \frac{\texttt{T}_{mcal}}{\texttt{T}_{PSF}}
\end{equation}

\noindent where $T_{mcal}$ and $T_{PSF}$ are sizes of the object or PSF respectively as defined in \NEW{Equation} \ref{eq:tsize}. \NEW{In this case the size is obtained from a single Gaussian fit, so the suffix \texttt{CM} (composite model) is not used.} Values are not constrained, but typical ranges explored for star-galaxy separation are between 0 and 1.

\subsection{ADA\_PROB}  
This is the name given to a machine learning framework using the {\tt scikit-learn}  package \citep{scikit-learn}.
\paragraph*{\textit{Input data:}} This method  uses feature generation \NEW{(using various simple mathematical functions of various catalogue variables)}, and feature pre-selection \NEW{(selecting the most informative variables).}
\paragraph*{\textit{Method:}} The selected quantities are fed into several machine learning algorithms (including AdaBoost) which are drawn from {\tt scikit-learn} with an additional probability recalibration step. The details of the framework are described in detail in Appendix \ref{sec:ada_prob_desc}. Two variants have been used of this approach, using either {\tt SExtractor} quantities {\tt ADA\_PROB} or {\tt MOF} quantities {\tt ADA\_PROB\_MOF}.

\subsection{GALSIFT\_PROB} 
A probabilistic estimate based on machine learning approach over principal components, as used in the `Multi\_class' algorithm in \citet{soumagnac}. 

\paragraph*{\textit{Input data:}} A principal component analysis (PCA) \NEW{over the catalogue quantities} is performed to outline the correlations between the object parameters and extract the most relevant information. We perform a calculation of the Fisher discriminant \citep{fisher} for each of the new parameters to quantify their aptitude to separate between the classes. 
\begin{equation}
\mathcal{F}_i = \frac{(\overline{X_{G,i}}-\overline{X_{S,i}})^2}{\sigma^2_{G,i}+\sigma^2_{S,i}}
\end{equation}
$G$ and $S$ corresponding to the galaxy and star classes respectively.

\paragraph*{\textit{Method:}} 
We select the parameters with the highest Fisher discriminant (hence the highest `separation power' of the classes) and use them as input to a machine learning classification algorithm. Whereas in \citet{soumagnac} the authors used {\tt ANNz} \citep{collister}, in this application we have replaced it by a Random Forest classification algorithm implemented as part of the \texttt{scikit-learn} package for Python \citep{scikit-learn}. The output is a probability of the object being a star or a galaxy. In this case, we have used a classifier based only on  {\tt MOF} quantities, {\tt GALSIFT\_PROB\_MOF}.

\subsection{SVM} 
Following Wei et al. (in prep), the support vector machine ({\tt SVM}) is a single-band, purely morphological and magnitude based classifier.
\paragraph*{\textit{Input data:}} The input features used by the SVM are \texttt{MAG\_AUTO\_I}, \texttt{FLUX\_RADIUS\_I}, and \texttt{SPREAD\_MODEL\_I}.
\paragraph*{\textit{Method:}} {\tt SVM} is a supervised machine learning algorithm that constructs a separating hyperplane in any arbitrary n-dimensional space that maximises the margins of objects to the hyperplane. To make the {\tt SVM} robust across various data sets with intrinsic variations in observation conditions, the algorithm performs linear transformations on the three input features to remove the means and make the standard deviations across all objects to be one. This preprocessing procedure also allows all three features to have equal levels of feature importance. This prevents any features with particularly large numerical values from dominating the {\tt SVM} classification decision. The {\tt SVM} uses a Gaussian radial basis function (rbf) kernel, where the hyperparameters, $\gamma = 0.01$ and $C = 46.4$, are selected while training the SVM through an exhaustive cross-validated grid search. The SVM outputs distances of objects to the hyperplane, where a high positive (negative) value corresponds to a high confidence star (galaxy) classification.


\subsection{CONCENTRATION} 
A parameter similar to what was used as a star-galaxy classifier for SDSS \citep{sdssDR2}\textcolor{red}. 
\paragraph*{\textit{Input data:}} The PSF and model magnitudes for each object.
\paragraph*{\textit{Method:}} In the case of DES, this translates to the use of the difference between the \texttt{MOF} PSF magnitude and a bulge + disk, or composite, model magnitude computed by the \texttt{MOF} pipeline:

\begin{equation}
\texttt{CONCENTRATION} = ${\tt PSF\_MAG\_I} - {\tt CM\_MAG\_I}$
\end{equation}

\subsection{W1-J, J-K infrared bands}
In the Stripe 82 region, we will compare with the information provided by the Vista Hemisphere Survey DR3 \citep{vhs} as proposed in \citet{banerji} up to the available depth. We will also estimate the classification power of a cut in the infrared bands from WISE \citep{wise}, 2MASS \citep{2mass}, as described in \citet{kovacs}. 
\paragraph*{\textit{Input data:}} Magnitudes W1 (WISE), J (2MASS, VHS) and K (VHS). 
\paragraph*{\textit{Method:}} Colour cuts in W1-J and J-K.

\COMMENT{\item \textbf{HB\_PROB} makes use of several templates for stars and galaxies (as developed and described in \citet{kim}), and uses a Hierarchical Bayesian approach to the estimation of the probability from the fluxes at different bands (\citet{fadely,kim}). A posterior probability is calculated using the standard expression (for a given class $A$):

\begin{equation}
P(A|\mathbf{x,\theta}) = P(\mathbf{x}|A,\mathbf{\theta})P(A|\mathbf{\theta})
\end{equation}

where the likelihood $P(\mathbf{x}|A,\mathbf{\theta})$ is computed by marginalizing over all star and galaxy templates. This requires specifying a prior probability $P(t|A,\mathbf{\theta})$ for every template $t$ that is obtained from the complete application sample itself.

The templates used follow the application described in \citet{kim} and include the stellar SEDs from \citet{pickles,chabrier,bohlin} and galaxy spectra from \citet{coleman,kinney}. 
}

\subsection{HB\_PROB}
Additionally, we implemented a Hierarchical Bayesian method (\texttt{HB\_PROB}) developed and explored by \citet{fadely,kim} with CFHTLS data. The lack of $u$-band in our case severely impacted the performance of this method, so it was not pursued further in our analysis.

Table \ref{tab:cuts} shows the specific selection methods used with respect to a varying threshold $t$ for each of the algorithms used in this work.

\begin{table*}
	\centering
	\caption{Selection methods.}
	\label{tab:cuts}
	\begin{tabular}{cc} 
		\hline
		Name & Selection method for galaxies using threshold \textit{t} \\
		\hline
		CLASS\_STAR & $CLASS\_STAR < t$ \\
		SPREAD\_MODEL & $SPREAD\_MODEL + 1.67*SPREADERR\_MODEL > t$\\
        CM\_T & $CM\_T + 2*CM\_T\_ERR > t$ \\
        MCAL\_RATIO & $MCAL\_RATIO > t$ \\
        ADA\_PROB & $ADA\_PROB > t$ \\
        GALSIFT\_PROB &  $GALSIFT\_PROB > t$ \\
        SVM & $SVM\_PROB > t$ \\
        CONCENTRATION & $PSF\_MAG\_I - CM\_MAG\_I > t$ \\
        WISE J-K & $(J-K-0.6)/(MAG\_AUTO\_G-MAG\_AUTO\_I) > t$ \\
		\hline
	\end{tabular}
\end{table*}

\section{Performance on calibration fields}
\label{sec:calib}

\COMMENT{We study the performance of the classifiers by separating point-like versus extended sources. The former will include both stars and QSOs on first approximation and we will collectively call them `stars' in this work. For the case studies considered here and the general catalogue, the contamination of QSOs in the large-scale stellar or galactic catalogues is not deemed important. However, a good star-QSO separation is needed for quasar science, as studied in detail in \citet{tiess} for DES data.QSO
 and analyze the impact in two broad science cases, and possible developments to improve object classification in future analyses of this data.}
 
In this section, we will look first at the metrics used to compare classifiers using the calibration fields, describe the datasets (including training and validation) and finally analyze the results.

\subsection{Receiver Operating Characteristic (ROC) curves}
We compare the performance of the different classification techniques using the calibration fields, by calculating Receiver Operating Characteristic \citep[ROC;][]{rocprimer,bradley} curves which compare the \textit{True Positive Rate (TPR)} of galaxy or star detection, given a specific threshold for the classifier, versus the \textit{False Positive Rate (FPR)}, as defined by:

\begin{equation}
    TPR=\frac{TP}{TP+FN}
	\label{eq:tpr}
\end{equation}
\begin{equation}
    FPR=\frac{FP}{FP+TN}
	\label{eq:fpr}
\end{equation}

\noindent where $TP$ are correctly identified galaxies, given a cut for a specific classifier; $FN$ are incorrectly classified galaxies as stars; $FP$ are incorrectly classified stars as galaxies and $TN$ correctly identified stars (in the assumption of using `truth' for galaxy type). See Table \ref{tab:confusion_matrix} for a reference on these concepts. Therefore, the ROC curve is confined by construction to an area spanning from 0 to 1 in FPR and TPR. As we vary the threshold $t$ for classification for a given classifier (Table \ref{tab:cuts}), a curve will be drawn across the area from (0,0) to (1,1). A completely random `classifier' would show as a diagonal line.

\begin{table*}
\centering
\caption{Definitions of different figures of merit for classifiers, according to the outcome of the classification using a `truth' reference (also termed `confusion matrix'). The term `positive` can refer to `galaxy' or `star' classes depending on the use case. The metrics examined in this work are emphasised in bold. \NEW{`Purity' can be used as a synonym for the Positive Predictive Value, PPV, whereas `completeness' can be interchanged with the True Positive Rate, TPR.}}
\label{tab:confusion_matrix}
\begin{tabular}{l|l||c|c||c|c|}
\multicolumn{2}{c}{} & \multicolumn{2}{c}{\textbf{Prediction}}& \multicolumn{2}{c}{} \\
\cline{3-4}
\multicolumn{2}{c|}{} & Positive & \multicolumn{1}{c|}{Negative} & \multicolumn{2}{c}{}\\
\hhline{~-::==:t:--}
\multirow{2}{*}{\textbf{Truth}}& Positive & \makecell{True Positive \\ (TP)} & \makecell{False Negative \\ (FN)} & \makecell{\textbf{True positive rate} \\ (TPR) = TP/(TP+FN)} & \makecell{False negative rate \\ (FNR) = FN/(TP+FN)} \\
\hhline{~-||--||--}
& Negative & \makecell{False Positive \\ (FP)} & \makecell{True Negative \\ (TN)} & \makecell{\textbf{False positive rate} \\ (FPR) = FP/(FP+TN)} & \makecell{True negative rate \\ (TNR) = TN/(FP+TN)} \\
\hhline{~-||~~||--}
\hhline{~~::==:b:~~}
\multicolumn{2}{c}{} & \multicolumn{1}{|c|}{\makecell{\textbf{Positive predictive value} \\ (PPV) = TP/(TP+FP)}} & \multicolumn{1}{|c|}{\makecell{False omission rate \\ (FOR) = FN/(FN+TN)}} & \multicolumn{2}{c}{}\\
\cline{3-4}
\multicolumn{2}{c}{} & \multicolumn{1}{|c|}{\makecell{False discovery rate \\ (FDR) = FP/(TP+FP)}} & \multicolumn{1}{|c|}{ \makecell{Negative predictive value \\ (NPV) = TN/(FN+TN)}} & \multicolumn{2}{c}{}\\
\cline{3-4}
\end{tabular}
\end{table*}

In particular, the AUC (area under the ROC curve) has been classically used as a threshold-independent metric to compare the performance of classifiers, as well as being relatively insensitive to the specific positive to negative composition (as long as sufficient statistics are available). The closer the AUC gets to unity, the better the discriminating power of the classifier associated with that particular curve. Again, a random classifier would show an AUC value of 0.5.

There are, however, some caveats to be aware of, namely the possibility of misleading results when ROC curves cross each other \citep{hand2009} and that misclassification costs can be different according to the scientific case, and this is not reflected in ROC curves. We address this by extending the range of metrics used for different classifiers, in order to have a broader view of the performance for our particular needs.

\subsection{Purity and completeness}
\label{sec:purity_completeness}
In astronomy, we are interested in evaluating the performance of classifiers in terms of their impact on measurable on parameters of interest. It is common to find the requirements for a survey defined in terms of \textit{purity} and \textit{completeness}. In \citet{soumagnac}, for example, the authors formulate the scientific requirements for weak lensing and large-scale structure studies in terms of these two observables.

`Purity' is a measurement of the contamination of a sample by misclassified objects, which can also be called \textit{precision} or \textit{positive predictive value (PPV)}:

\begin{equation}
    PPV=\frac{TP}{TP+FP}
	\label{eq:ppv}
\end{equation}

`Completeness' (also known as, \textit{recall}) is another name for the TPR defined in Equation \ref{eq:tpr}. A good approach to easily compare the performances of several classifiers is to use the precision-recall (PR) curve, where both quantities can be visualised simultaneously.

\subsection{Training and testing fields}
\label{sec:train_test_fields}

The dataset on which we train the machine learning (ML) codes is the weak lensing catalogue from HST ACS in the COSMOS field\citep{leauthaud}, as this provides a largely unbiased measurement of all extended and point-like sources from DES (albeit the star-galaxy mixture is affected by the specific position in the sky with respect to the Galactic plane). In particular, the \texttt{MU\_CLASS} parameter is used for this reference, defined in the peak surface brightness - \texttt{MAG\_AUTO} space, which in space-based imaging shows very distinct loci with respect to the same objects viewed through the atmosphere. This has been used previously in star-galaxy separation assessments in, e.g., \citet{crocce} and \citet{hypersuprime}.

This training set, after a 1" positional match with DES sources, contains $\sim114$k extended and $\sim12$k point-like sources. \COMMENT{It can be subdivided further into a training set and a validation set, the latter being used to tune each classifier's internal parameters.} The COSMOS dataset will also be used for some tests only with the non-ML codes in order to avoid biased conclusions based on their training in that same area. 

Even in the case in which we use unbiased, imaging data, the particular position on the sky of the field will condition the relative mixture of stars and galaxies in a prominent way. Therefore we add some extra imaging data extracted from the Hubble Source Catalog\footnote{\url{https://archive.stsci.edu/hst/hsc/}} (Hubble-SC) \citep{hsc} where it overlaps the DES survey. Most of it is either too inhomogeneous or targets specific objects (nearby, large galaxies or globular clusters), but a few deep fields can be matched with some of the SN fields from DES. In this case we use the Hubble-SC catalogues' concentration index with a cut of 1.2 which seems optimal in the concentration-magnitude plane.


Spectroscopy is also a valuable resource to provide a one-to-one truth table for our classifications. However, the spectroscopic targeting and measurement efficiency is not complete in a statistical sense relative to the DES catalogue, as certain types of sources were given higher priority and some types are more difficult to classify spectroscopically,  therefore the testing of purity/completeness can be strongly biased. The photometric properties of the stars and galaxies selected can also be highly skewed to particular types that introduce additional biases. This limits the usefulness of any purity metric we try to derive from these fields. For this reason, the spectroscopic datasets have been limited to those that provide a relatively unbiased sample by construction, which includes the VVDS-DEEP and VVDS-CDFS \citep{vvds} data releases. The SDSS DR13 \citep{sdssdr13} updated spectro-photometric sample over Stripe 82 is also used due to the relative variety of spectra available, and the possibility to test our classification methods against `true' spectroscopic typing. We use redshifts (a cut in z < 0.001) as the method to identify stars. However, we also consider a selection based on SDSS spectroscopic {\tt CLASS} \NEW{obtaining similar conclusions. For the VVDS data, we require the redshifts to be `reliable' according to the classification in \citet{vvds}, that is, values of 2,3,4 or 9 in their redshift quality estimate.}

Both the COSMOS catalogues and the ones recovered from the Hubble-SC have been cross-tested against spectroscopic catalogues (VIMOS-Ultra Deep Survey DR1 \citep{vuds}, zCOSMOS DR3 \citep{zcosmos}, and VVDS-CDFS \citep{vvds}) to check the robustness of their morphological classifications against a `true' type based on their spectra. In both cases, around $5\%$ of spectroscopically classified stars are misclassified as galaxies \NEW{when using these space-imaging based measurements}, whereas around $2\%$ of spectroscopically classified galaxies are misclassified as stars. This \NEW{misclassification happens at faint magnitudes (F814W from ACS-HST > 24 for COSMOS, F814W > 23 for the other Hubble fields used), denoting possible compact galaxies that are unresolvable by HST, errors in the spectroscopic measurement or matching.} These corrections are not considered for the purity estimates derived here as they belong to fainter fluxes than the truth tables used in our tests. 

See Table \ref{tab:external_datasets} and Appendix \ref{sec:external_datasets} for details on the reference data in different fields including the database queries used to create these datasets.

\subsection{Results}

\subsubsection{Using HST imaging}
\label{sec:imaging_results}

We compare here the results for the classifiers used on the COSMOS field (excluding the ML codes that were trained on this field) and the supernova fields for which we have found publicly available deep HST data from the Hubble-SC.

\begin{figure}
	\includegraphics[width=\columnwidth]{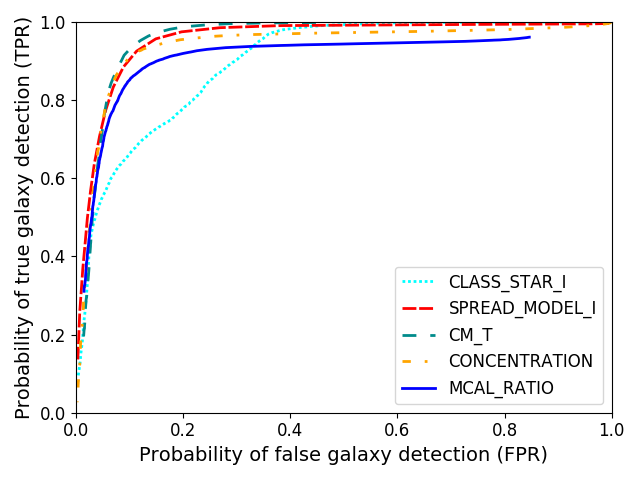}
    \caption{ROC plot for classifiers tested on the COSMOS field. Only non-ML codes are shown, as  \NEW{the machine-learning ones were trained in this dataset}. Magnitude range is given by \texttt{MAG\_AUTO\_I} = (17,24). The {\tt SPREAD\_MODEL}-based cut is similar to {\tt MODEST\_CLASS} used in Y1 analyses. \NEW{The ROC curve is obtained by varying the threshold at which the classification divides the galaxy and star sample.}}
\label{fig:roc_cosmos}
\end{figure}

\begin{figure}
	\includegraphics[width=\columnwidth]{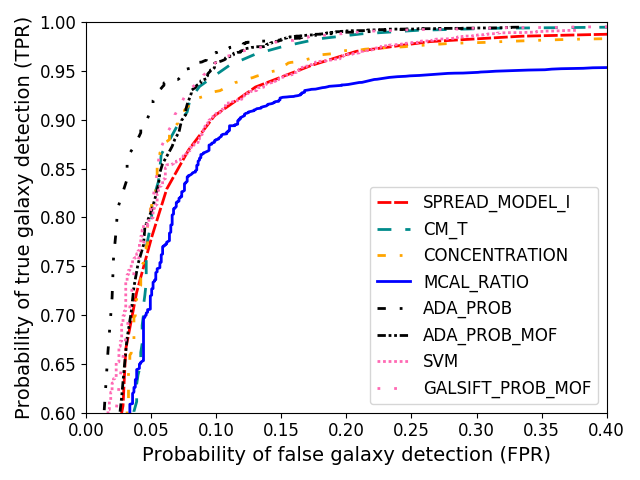}
    \caption{ROC plot for classifiers tested on the SN fields over the Hubble-SC catalogue. Magnitude range is given by \texttt{MAG\_AUTO\_I} = (17,24). The {\tt SPREAD\_MODEL}-based cut is similar to {\tt MODEST\_CLASS} used in Y1 analyses. \NEW{The ROC curve is obtained by varying the threshold at which the classification divides the galaxy and star sample.}}
    \label{fig:roc_snfields}
\end{figure}

\begin{figure}
	\includegraphics[width=\columnwidth]{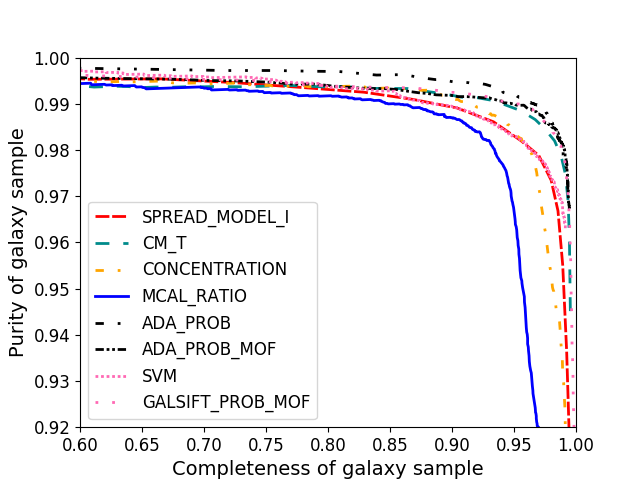}
    \caption{Precision-Recall (or completeness-purity) plot for classifiers tested on the SN fields over the Hubble-SC catalogue, using \textbf{galaxies} as truth. Magnitude range is given by \texttt{MAG\_AUTO\_I} = (17,24). The {\tt SPREAD\_MODEL}-based cut is similar to the {\tt MODEST\_CLASS} used in DES Y1 analyses.}
    \label{fig:pr_snfields}
\end{figure}

\begin{figure}
	\includegraphics[width=\columnwidth]{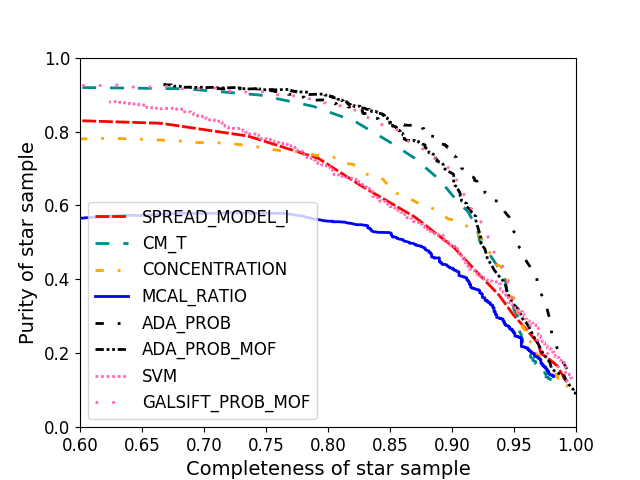}
    \caption{Precision-Recall (or completeness-purity) plot for classifiers tested on the SN fields over the Hubble-SC catalogue, using \textbf{stars} as truth. Magnitude range is given by \texttt{MAG\_AUTO\_I} = (17,24). The {\tt SPREAD\_MODEL}-based cut is similar to the {\tt MODEST\_CLASS} used in DES Y1 analyses.}
    \label{fig:pr_snfields_stars}
\end{figure}

\begin{figure}
	\includegraphics[width=\columnwidth]{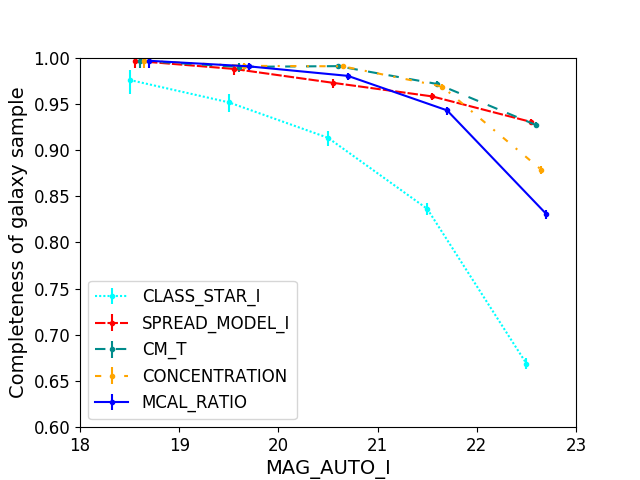}
    \caption{Completeness of a {\bf galaxy} sample as a function of magnitude for classifiers tested on the COSMOS field, for a fixed galaxy purity of 99\%.}
    \label{fig:efficiency_mag_cosmos}
\end{figure}

\begin{figure}
	\includegraphics[width=\columnwidth]{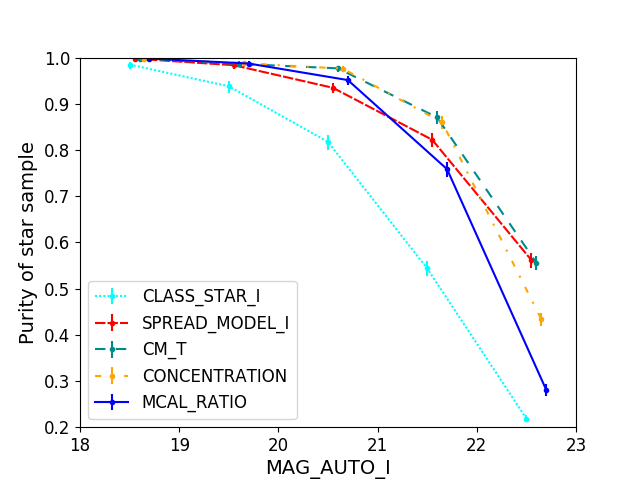}
    \caption{Completeness of {\bf stellar} sample as a function of magnitude for classifiers tested on the COSMOS field, for a fixed 80\% purity.}
    \label{fig:efficiency_mag_cosmos_stars}
\end{figure}

\begin{itemize}
\item The results for the ROC comparison are shown in Figure \ref{fig:roc_cosmos} for the COSMOS field and  Figure \ref{fig:roc_snfields} for the SN fields with Hubble-SC data. The AUC of the respective curves are tabulated in Table \ref{tab:aucs}. 

From these plots it can be readily seen that among the morphological classifiers, the algorithms based on a linear discriminant over coadded images, {\tt SPREAD\_MODEL}, and intrinsic size on {\tt MOF} estimates, {\tt CM\_T}, are the best performing ones. 
It is also seen that the ML classifiers (in Figures \ref{fig:roc_snfields} and \ref{fig:pr_snfields}) do perform better, even considering a different field with respect to training as in the case of the Hubble-SC test\COMMENT{, though admittedly the truth samples have been collected using the same instrument (the ACS camera with the F814W filter) in both training and testing (though the objects in these samples are different)}. 
It is noteworthy to point out that most of the differences showcased in Figures \ref{fig:roc_cosmos} and \ref{fig:roc_snfields} become more evident when we restrict ourselves to faint objects ($i>22$). The {\tt SPREAD\_MODEL}-based cut does a good job at avoiding stellar contamination but suffers from decreased galaxy completeness. This is a result of the galaxy locus merging with the stellar locus in the magnitude-{\tt SPREAD\_MODEL} space where noisier measurements will increase the effect even further. {\tt CM\_T} fares better in this respect, but a conservative cut will provide a more pure galaxy sample using {\tt SPREAD\_MODEL}. On the other hand, the metacalibration size ratio does not perform as well as the other morphological classifiers, though this measurement is noisier than the direct assessment of sizes and shapes from the {\tt MOF} pipeline.  
\item Figure \ref{fig:roc_snfields} shows that ML classifiers are able to take advantage of ancillary information for very faint objects where shape measurements are uncertain. Results with {\tt SVM} in the SN fields show that a ML approach based exclusively on morphological and magnitude information can provide some advantage over simple cuts on morphological variables. {\tt SVM} is shown to be robust outside of its training field, however other machine learning algorithms provide an extra edge in performance as shown by the higher AUC values. This is due to forgoing the additional information encoded in the rest of the variables available in the catalogue. However, this approach could provide a middle-ground solution to the issues one might encounter when incorporating colour-based information, which can incorporate interesting physics we would not like to be entangled with our star-galaxy sample selection (see Section \ref{sec:discussion}). Further developments of this approach is explored in Wei et al. (in prep).

The comparison between the COSMOS and Hubble-SC fields reveals that the {\tt CM\_T} classification is more robust as we switch between fields. {\tt SPREAD\_MODEL} and {\tt CLASS\_STAR}, which are derived from coadded PSFs are more vulnerable to the contribution of bad exposures and PSF inhomogeneities in the coadded image. It is worthwhile noting here that preliminary tests on Y3 data \citep{dr1} using Hyper Suprime Camera deep data \citep{hypersuprime} reinforce this idea, which will be explored further in a future publication, therefore favoring in general the use of a multi-epoch classifier (such as {\tt CM\_T} based on the {\tt MOF} pipeline). Both the COSMOS field dataset and SN field coadds have a much smaller dithering than the wide-field exposures. This might artificially bias classifications based on the coadded PSF to somewhat better performances than actually present in the wide-field data. 

\item Figures \ref{fig:pr_snfields} and \ref{fig:pr_snfields_stars} show the precision-recall metric, for galaxies and stars respectively (COSMOS plots not shown for conciseness, but provide similar conclusions).

These plots provide a similar conclusion as the ROC curves, though in terms of more useful quantities with respect to scientific requirements such as the recall (i.e. completeness) and precision (i.e. purity). Again, the {\tt CM\_T} morphological classifier and the ML codes provide the best results, and this manifests even more strongly for selecting a star sample (these results motivate the choice for stellar classification based on multi-epoch pipelines in \citealt{shipp}). It is noteworthy to add that the ML classifiers using {\tt MOF} quantities do not add much more than a straight cut in {\tt CM\_T} itself, due to the large information content included in this classifier with regards to star-galaxy classification. On the other hand, the ML classifiers based on \texttt{SExtractor} quantities are able to extract more value from the different outputs of this code, with respect to a simple {\tt SPREAD\_MODEL} cut.

\item In Figures \ref{fig:efficiency_mag_cosmos} and \ref{fig:efficiency_mag_cosmos_stars}, we can appreciate the dependence of the completeness with the magnitude as we go to the fainter end in the sample, in the galactic and stellar case respectively.

Unlike in the previous plots, in this case a choice of threshold has to be made. We have decided to pick cuts in the variables in question in order to have a similar galaxy purity ($99\%$) in each magnitude bin, so we can compare completeness appropriately, and similarly for stars ($80\%$). We chose the COSMOS field which has good statistics to faint magnitudes, though this disallows using the ML codes in the comparison. This example shows a case where classifiers such as the concentration estimation from the {\tt MOF} pipeline, not necessarily favored at first sight from the integral under the ROC curve, works better in this regime due to its good selection of very pure samples. The ROC curve only informs about \textit{overall} classifier performance (i.e. considering all possible thresholds), and different classifiers have to be tested for the specific science case at hand. 

For stars, a similar behaviour is seen for {\tt CM\_T}, {\tt CONCENTRATION} and {\tt SPREAD\_MODEL}. {\tt CLASS\_STAR} for instance suffers from a poor completeness near the faint end, as a high thresholding cut in this case removes most of the objects, which in the neural network tend to cluster towards intermediate values when the object classification is uncertain. {\tt MCAL\_RATIO} incorporates noisier measurements and additional cuts to the sample that make it less complete when providing a classified sample.

\item In addition, in Figure \ref{fig:purity_pz_hsc} a similar comparison is shown as a function of a realization of the photometric redshift from the probability distribution function obtained from the algorithm BPZ \citep{bpz}, this time also adding the ML classifiers (again over the SN fields with Hubble-SC). A similar conclusion is drawn from these plots; {\tt MOF} fitting methods and ML classifiers perform best, as indicated by the ROC curves. Note the stability of the purity of the galaxy sample with respect to photo-z, suggesting that a photo-z selected sample would not be biased by the star-galaxy separation classifiers analyzed here (however, see Section \ref{sec:lss} for an important caveat to this conclusion). 
\end{itemize}

\begin{table*}
	\centering
	\caption{Area under the ROC curves for different classifiers. Dashes indicate tests that have not been run for that specific code and dataset combination.}
	\label{tab:aucs}
	\begin{tabular}{ccccc} 
		\hline
		Name & COSMOS, imaging & SN fields, imaging & SN fields, spectroscopy & stripe 82, spectroscopy\\
		\hline
		CLASS\_STAR & 0.898 & 0.885 & 0.950 & \NEW{0.976} \\
		SPREAD\_MODEL & 0.954 & 0.956 & 0.975 & \NEW{0.962} \\
        CM\_T (MOF) & 0.957 & 0.959 & 0.971 & \NEW{0.972} \\
        CONCENTRATION (MOF) & 0.938 & \NEW{0.953} & 0.950 & \NEW{0.967} \\
        MCAL\_RATIO & 0.910 & 0.924 & -- & -- \\
        VHS J-K vs G-I & -- & -- & -- & \NEW{0.993} \\
        ADA\_PROB & -- & 0.978 & 0.983 & \NEW{0.967} \\
        ADA\_PROB (MOF) & -- & \NEW{0.967} & 0.980 & \NEW{0.967} \\
        GALSIFT\_PROB (MOF) & -- & \NEW{0.969} & 0.981 & \NEW{0.962} \\
        SVM & -- & 0.962 & -- & -- \\
		\hline
	\end{tabular}
\end{table*}

\begin{figure}
	\includegraphics[width=\columnwidth]{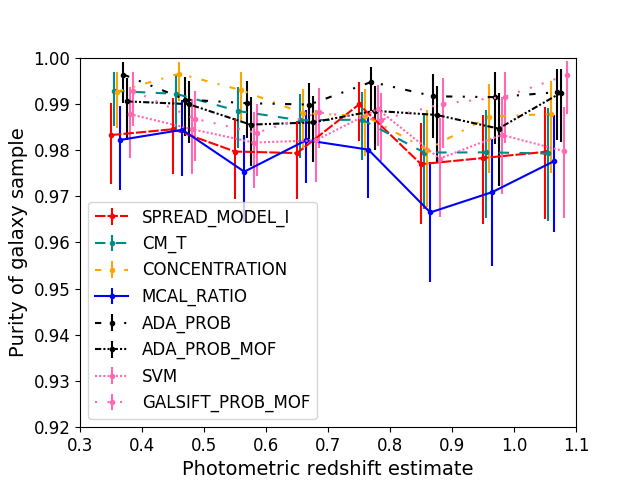}
    \caption{Purity of the galaxy sample as a function of photo-z for classifiers tested on Hubble-SC matches over the SN fields field, for a fixed 90\% completeness. We use a random MonteCarlo sampling of the probability distribution function of redshift predicted by BPZ for that particular object as an estimate of its photo-z.}
    \label{fig:purity_pz_hsc}
\end{figure}






\subsubsection{Using ground-based spectroscopy}
\label{sec:spectralcalib}

Turning now to tests on the overlapping spectroscopic data, we show ROC plots to demonstrate the consistency with the results from the previous section and add a comparison with external infrared information. 

\begin{figure}
	\includegraphics[width=\columnwidth]{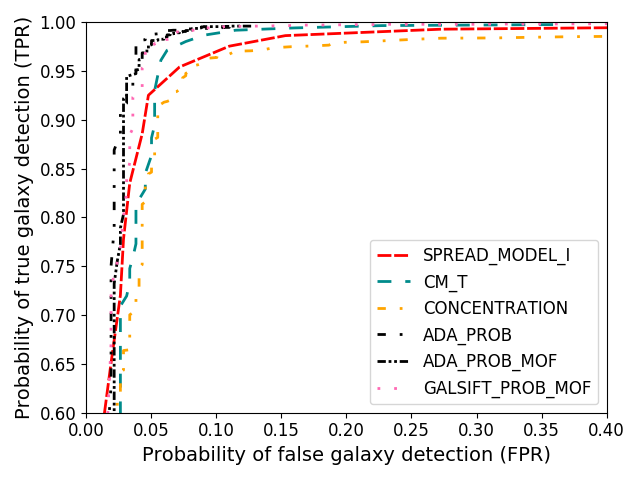}
    \caption{ROC plot for classifiers tested on the SN fields over the VVDS catalogues. Magnitude range is given by \texttt{MAG\_AUTO\_I} = (17,24). \NEW{The ROC curve is obtained by varying the threshold at which the classification divides the galaxy and star sample.}}
    \label{fig:roc_vvds}
\end{figure}

\begin{figure}
	\includegraphics[width=\columnwidth]{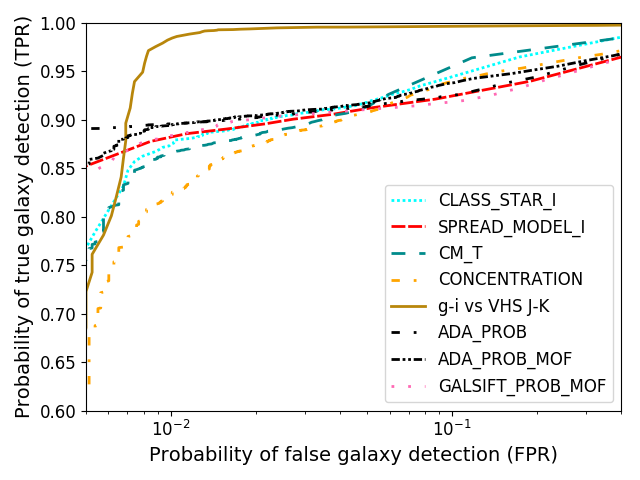}
    \caption{ROC plot for classifiers tested on the Stripe 82 region overlapping SDSS and VHS data. Magnitude range is given by \texttt{MAG\_AUTO\_I} = (17,21). Note the logarithmic scale in the x-axis in this instance. \NEW{The ROC curve is obtained by varying the threshold at which the classification divides the galaxy and star sample.}}
    \label{fig:roc_s82}
\end{figure}

\begin{figure}
	\includegraphics[width=\columnwidth]{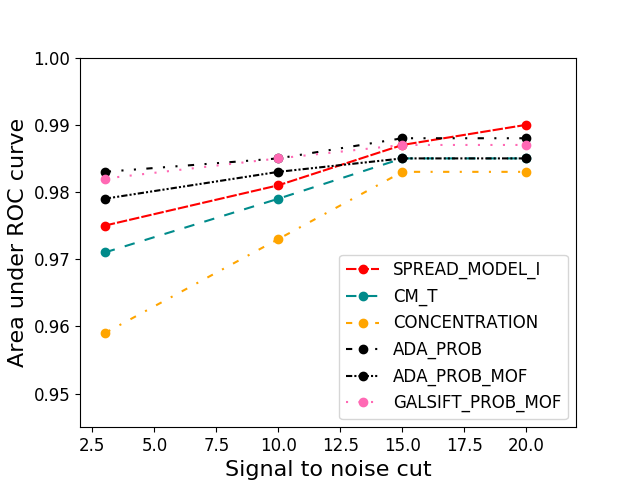}
    \caption{Area under the curve measured for the same classifiers as Figure \ref{fig:roc_vvds}, for different signal-to-noise thresholds, using the \texttt{MAGERR\_AUTO} quantity.}
    \label{fig:aucs_sn}
\end{figure}

Figure \ref{fig:roc_vvds} shows the ROC for the VVDS test and Figure \ref{fig:roc_s82} shows the ROC for the Stripe 82 test. The former does not add much to the conclusions mentioned above, but provides an assurance that conclusions are consistent with a different class of `truth' typing. We also add here a test on the SN fields computing the ROC curves and their areas, versus the signal to noise of the detected objects, to demonstrate it behaves as expected as well, including the ML codes (see Figure \ref{fig:aucs_sn}). 

The stripe 82 dataset is shallower and therefore does not allow for a clear distinction between the performance of most of the algorithms described here. The comparison with the combination with external infrared colour cuts on the other hand, shows an important increase in performance, specifically when attempting to select a very pure stellar sample, as already advanced in \citet{baldry} and \citet{banerji}. It is important to note here again that the nature of the test is different with respect to the ones based on space imaging. In this case we are using spectroscopic redshifts to determine the nature of the object (galactic or extra-galactic) and not its extendedness. What we see here is that infrared information will select out the stars from the galaxy and QSO (which are point-like generally) population. We have also attempted to add W1-J version from 2MASS and WISE (as suggested in \citet{kovacs}) but the matches proved too shallow to be of any interest for these samples. 

Unfortunately, the current VHS data does not cover the full breadth and depth of the survey and a careful combined catalogue with adequate matching is needed (overcoming the less precise infrared astrometry) beyond what was done here for comparison purposes. Cross-matching with bright sources will be explored in more detail with DES Y3 data with the goals of enhancing star selection for creating PSF models and reference catalogues for large scale structure. A combination of classifiers, as done for instance in \citet{kim} or \citet{molino}, seems to be an appropriate option in this case and even more so if \NEW{matched-aperture} photometry of VHS data can be performed survey-wide for DES \citep{banerji}. This would also have important applications for photometric redshift determination \citep{banerjipz}.



\section{Performance on application field}
\label{sec:y1valid}

It has been shown by \citet{fadely} that machine learning techniques in star-galaxy classification will perform better if a representative training dataset is found. We have studied the impact of this effect by testing ML algorithms over different fields other than the training set in Section \ref{sec:calib}. However all these additional areas are quite constrained either in depth or area, when compared to the complete DES volume.

In this section, we extend the scope of the performance tests in classification to have a broader picture, by making the following checks on the application field (see Section \ref{sec:dataset}):

\begin{enumerate}
\item General distribution of the classifier-flux space to qualitatively analyze the algorithms' outputs.
\item Number count distributions of stars against a well-tested simulation, both as a function of magnitude and as function of galactic latitude.
\item Galaxy versus star density profiles in search of correlations, using different proxies for the true stellar distribution.
\item Density of classified galaxies as a function of proximity to the Large Magellanic Cloud. 
\item Consistency of classified stars with the expected stellar locus \citep{covey}.
\end{enumerate}

Except where noted, the sample sizes for each of these cases are approximately 1 million objects, limited by the size of tested region, magnitude range or photo-z binning. 

\subsection{Classifier outputs}

A first step towards understanding the quality of classification for different algorithms in the application field of DES is to study the outputs as a function of magnitude and the number counts of classified objects. 

\begin{figure*}
	\centering
    \subfigure[CLASS\_STAR]{\includegraphics[width=0.3\textwidth]{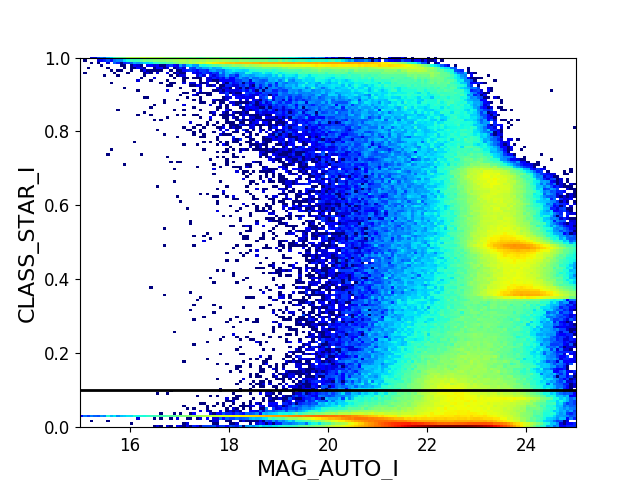}\label{fig:a}}
    \subfigure[SPREAD\_MODEL]{\includegraphics[width=0.3\textwidth]{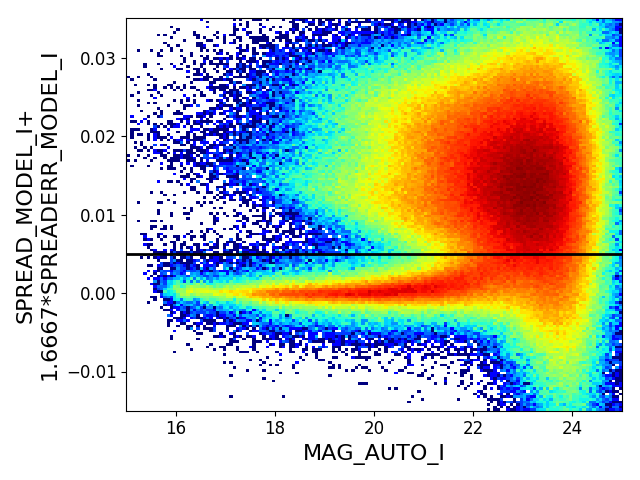}\label{fig:b}}
    \subfigure[CM\_T]{\includegraphics[width=0.3\textwidth]{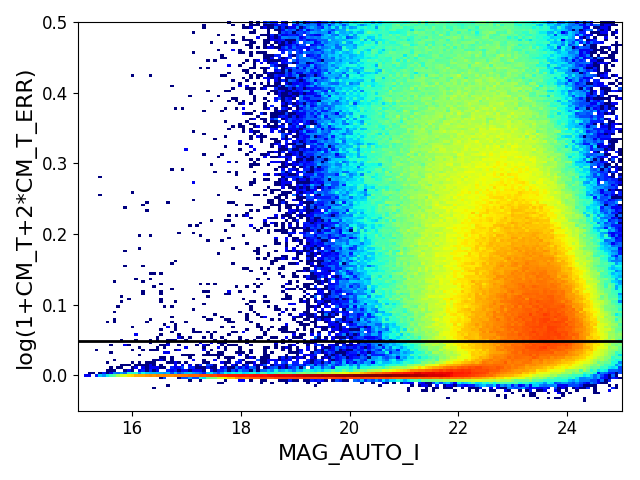}\label{fig:c}}
    \\
    \subfigure[ADA\_PROB]{\includegraphics[width=0.3\textwidth]{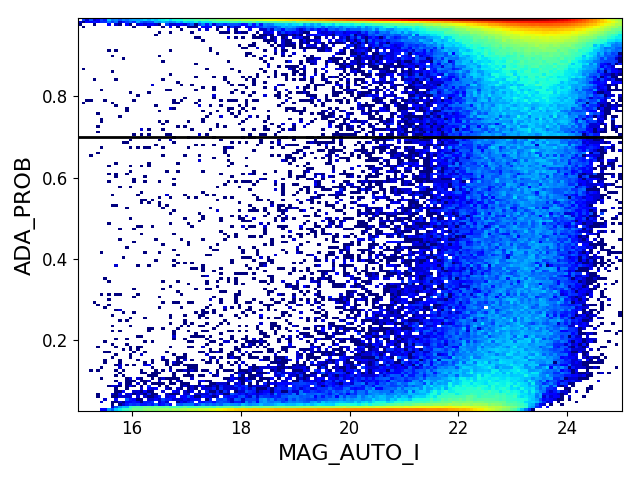}\label{fig:d}}
    \subfigure[ADA\_PROB\_MOF]{\includegraphics[width=0.3\textwidth]{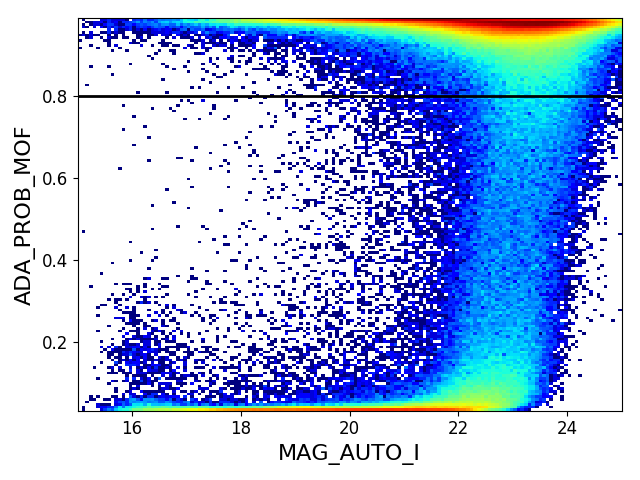}\label{fig:e}}
    \subfigure[GALSIFT\_PROB\_MOF]{\includegraphics[width=0.3\textwidth]{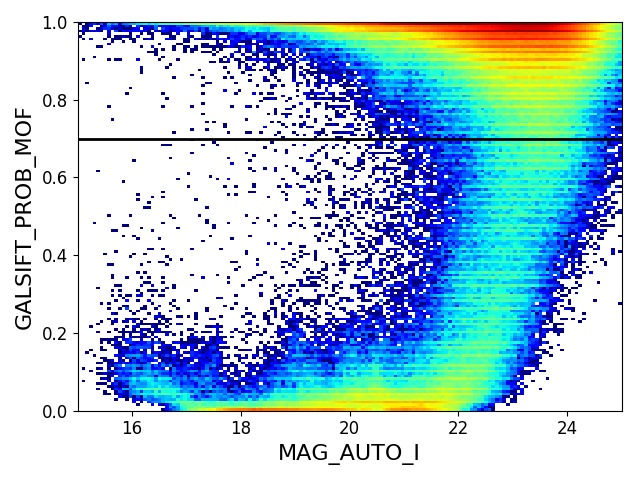}\label{fig:f}}
 	\caption{Object classification heatmaps as a function of magnitude for different classifiers. The black line represents the cut for which a 99\% galaxy purity is obtained in the Hubble-SC sample in the i=(17,24) magnitude range. With the exception of CLASS\_STAR, all classifiers assign higher values to extended sources.} \label{fig:magclassifier_distributions}
\end{figure*}

In Figure \ref{fig:magclassifier_distributions} several density plots showcase how objects distribute in the classifier-magnitude space. These distributions are based on a $1\%$ sample of the Y1 Gold catalogue. Direct morphological outputs from the DESDM pipeline {\tt CLASS\_STAR}, {\tt SPREAD\_MODEL} and {\tt CM\_T}) show two loci that merge in the faint end. {\tt CLASS\_STAR} outputs merge into a region of 50\% probability by construction of its base neural network. This uncertainty region appears at shallower magnitudes than other classifiers as shown previously, due to the characteristics of the simulations used for its training. \COMMENT{A similar effect is seen with \texttt{GALSIFT\_PROB}.} However, a classifier using a feature importance selection\footnote{A preselection of the input variables which provide the most predictive power for the task at hand, e.g. star-galaxy separation.} manifests a more `clear-cut' classification of objects, with a large predominance of galaxies at the faint end, as expected. This can be attributed to the fact that there is a large predominance of galaxies over stars in raw numbers (a very imbalanced dataset) at faint magnitudes, so the algorithms will `learn' that the most probable classification for a given object in this range is a galaxy.

\subsection{Number counts of classified stars}
\label{sec:starcounts}

\begin{figure}
	\centering
    \includegraphics[width=0.4\textwidth]{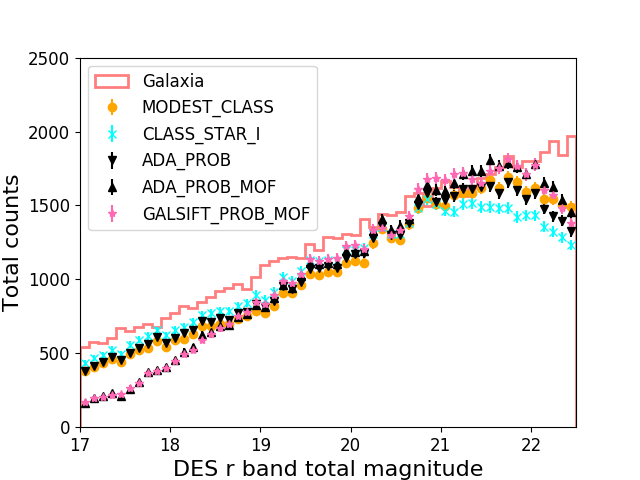}
   \caption{Counts for stars as classified by different algorithms compared to a {\tt Galaxia} simulation \citep{galaxia} using DES photometry, in the patch of the Y1 DES footprint with 45<RA<50, -45<DEC<-50.}\label{fig:nm_stars}
\end{figure}

On the other hand, if we limit our study to the point in which Y1 data are fairly complete over a large area ($r\sim22.5$), we can assess for instance the similarity of the stellar distribution in magnitude versus a detailed simulation such as {\tt Galaxia} \citep{galaxia}, which has been tested against Gaia DR1 data (\cite{gaiadr1}, Koposov private communication). This is shown in Figure \ref{fig:nm_stars} for a few selected classifiers, spanning a varied range of those mentioned in Section \ref{sec:classifiers}, in the DES \textit{r} band. Thresholds were used to provide a similar number of stars as {\tt MODEST\_CLASS}, the default DES Y1 Gold star-galaxy classifier based on {\tt SPREAD\_MODEL}. Up to $r\sim21$, the behaviour for most of them with respect to the simulation is similar. Two machine learning classifiers based on {\tt MOF} quantities show a significant lack of brigh objects ($r<19$) due to failures from the Y1 version of the {\tt MOF} pipeline in fitting stars in this regime\footnote{Y3 Gold {\tt MOF} photometry has solved this issue.}. This has been identified as failures of the galaxy fits for which {\tt MOF} was designed when applied to moderately bright stars. A consistent overestimation of stars by {\tt Galaxia} with respect to DES stars is apparent for all classifiers, as was seen in \citet{tingli}. On the other hand, other simulations such as the ones described in \citet{besancon} and \citet{trilegal} show discrepancies of this size as well at this latitude and longitude. 
This disappears at the faint end, as compact galaxies start to leak into the stellar sample. After that, a completeness drop kicks in as we enter the survey's magnitude limit. At the faint end, {\tt CLASS\_STAR} shows a drop in completeness sooner than the other classifiers. The nature of this classifier, which provides an intermediate value of probability for `uncertain' sources, is such that a fixed threshold cut tends to `lose' stars at the faint end, if we adjust all classifiers to the same number of stars. \COMMENT{For {\tt GALSIFT\_PROB} the effect comes from an anomalous behaviour at $i\sim21$, also seen in Figure \ref{fig:f} so that a more or less pure cut in the selection will make this kink appear. A somewhat looser cut will have the effect of smoothing these feature, though admitting a few galaxies in the faintest end.}

\subsection{Stellar density as a function of Galactic latitude}

\begin{figure}
	\centering
    \includegraphics[width=0.4\textwidth]{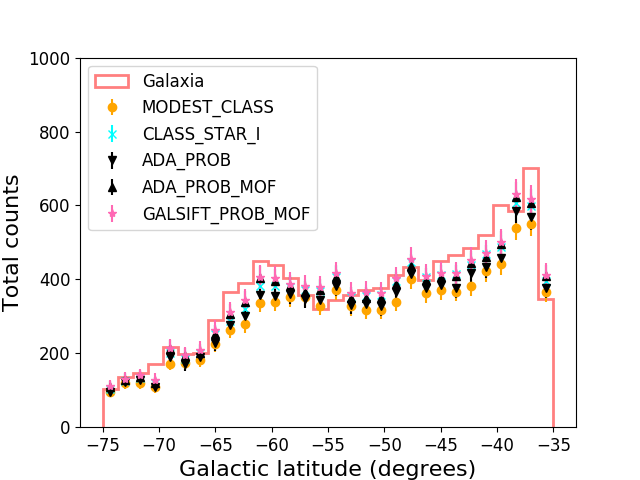}
   \caption{Counts for stars as classified by different algorithms compared to a Galaxia simulation \citep{galaxia} for the application field (SPT region of the DES-Y1 footprint) for the magnitude range $r$ = (19,21.5).}\label{fig:gallat_counts_stars}
\end{figure}

As a complementary measure of goodness of stellar identification, we compare the number of stars as a function of Galactic latitude (Figure \ref{fig:gallat_counts_stars}). We limit the comparison to the range in which any possible issues deriving from the current {\tt MOF} processing are avoided (see Section \ref{sec:starcounts}). A slight deficit is seen nonetheless as was verified before, but the comparison of all these different approaches are qualitatively in the same range, without any preferred or outstanding behaviour from any of the classifiers tested here.


\subsection{Galaxy vs stellar density}
\label{sec:densvsdens}

As mentioned in Section \ref{sec:purity_completeness}, we do not have a large-scale `truth' table available that we could use as reference to check the precision of our classification on an object-by-object basis. However, several studies of large-scale structure \citep[e.g.,][]{ross} have devised an estimate of the purity of the galaxy sample, for a given classification scheme, by measuring correlations of classified galaxy density versus some reliable measurement of the relative stellar distribution (using a very pure cut for stars, a model, or an external catalogue). This is done via the pixelisation of the field using the \texttt{HEALPix} software \citep{healpix} and fitting a linear relation between the galaxy overdensity as a function of stellar density in said pixels. For this study we used a pixelisation parameter \texttt{NSIDE}=512, which corresponds to a pixel size of approximately 0.01 square degrees.

\begin{figure}
	\includegraphics[width=0.45\textwidth]{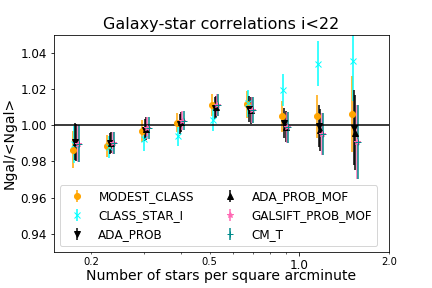}
	\includegraphics[width=0.45\textwidth]{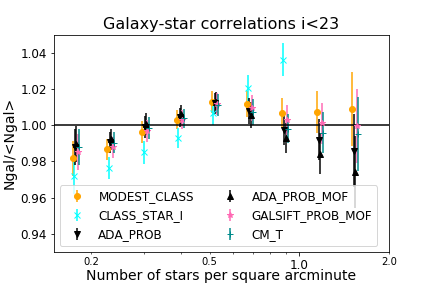}
    \caption{Galaxy vs star density plot for several classifiers, \NEW{for $i<22$ (\textit{top}) and $i<23$ (\textit{bottom}). Star density is traced by an external map of `secure' moderately bright stars.}}
    \label{fig:classcomp_densdens}
\end{figure}

In Figure \ref{fig:classcomp_densdens} we show a comparison \NEW{of the galaxy density as a function of stellar density for}  several classifiers, tested on the application field for the galaxy sample with the magnitude cuts shown in Table \ref{tab:contamination}. \NEW{Errors for each point are computed using the jackknife method \citep{jackknife}, whereas the ones in the table correspond to the estimated error from the fit.}

\NEW{The galaxy density over samples of increasing stellar density would theoretically increase with a linear relationship, if stellar contamination was the only effect that a dense star field would introduce. However, as seen already in \citet{ross} (in their Figure 3), moderately bright stars can also induce an `occultation' effect which makes detection around them more difficult. This effect is more predominant for fainter sources. This will create an inverse, possibly non-linear, relationship between galaxy density and stellar density. The overall effect is to create a proportionality relationship at low to moderate stellar densities, which may or may not change in slope and even decrease, depending on the separation power of the classifier, as galaxies get removed from the catalog due to the presence of foreground bright stars. For our purposes here, i.e., to understand the star-galaxy separation power for different classifiers, we use
the intercept value of the linear fit to the first part of the plot, in order to} estimate the purity of the galaxy sample. We adjusted the cuts for the classifiers to provide a similar number of detected `galaxies' (i.e. a similar completeness) as {\tt MODEST\_CLASS}, in order to get a better handle on how purity compares on the same grounds, similarly to what we did on Section \ref{sec:imaging_results}. 

We note that using the application sample in bulk shows no strong contamination component for the {\tt SPREAD\_MODEL}- or \texttt{MOF}-based quantities or for the machine learning approaches using magnitude and colour information. Slightly better performance is found using \texttt{MOF} quantities and the \texttt{ADABOOST} code, especially for fainter objects.
This is explained by the more accurate shape measurement of the \texttt{MOF} code and by how additional information is captured by {\tt ADA\_PROB\_MOF}.

\begin{table*}
	\centering
	\caption{Contamination for different classification methods for the galaxy vs stellar density tests. Threshold cuts were selected to adjust to the same number of detected galaxies as provided by {\tt MODEST\_CLASS}.}
	\label{tab:contamination}
	\begin{tabular}{ccccccc} 
		\hline
		Sample & MODEST\_CLASS & CLASS\_STAR & ADA\_PROB & ADA\_PROB\_MOF & GALSIFT\_PROB\_MOF & CM\_T\\
		\hline
		i$<22$ & \NEW{$2.7\pm0.4\%$} & \NEW{$2.1\pm0.5\%$} & \NEW{$2.2\pm0.4\%$} & \NEW{$2.2\pm0.4\%$} & \NEW{$2.3\pm0.4\%$} & \NEW{$2.3\pm0.4\%$}\\
        i$<23$ & \NEW{$3.2\pm0.4\%$} & \NEW{$4.6\pm0.2\%$} & \NEW{$2.4\pm0.4\%$} & \NEW{$2.1\pm0.4\%$} & \NEW{$2.8\pm0.3\%$} & \NEW{$2.4\pm0.4\%$}\\
		\hline       
	\end{tabular}
\COMMENT{	\begin{tabular}{cccc} 
		\hline
		Sample & GALSIFT\_PROB & GALSIFT\_PROB\_MOF & CM\_T\\
		\hline
		i$<22$ & $2.0\pm0.3\%$ & $0.8\pm0.6\%$ & $0.8\pm0.6\%$\\
        i$<23$ & $1.8\pm0.4\%$ & $1.2\pm0.6\%$ & $0.8\pm0.6\%$\\
		\hline       
	\end{tabular}}
\end{table*}

\begin{figure}
	\includegraphics[width=\columnwidth]{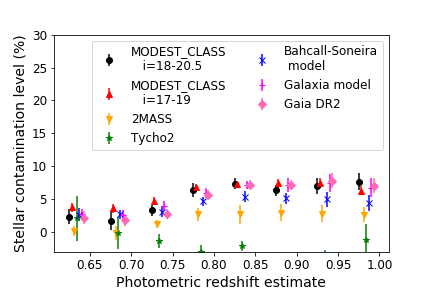}
    \caption{Star contamination levels for different stellar maps. A $\sim1\%$ systematic uncertainty is derived by comparing the {\tt MODEST\_CLASS} moderate to bright stars, is estimated from this plot. Tycho and 2MASS stars are added for comparison, but their magnitude ranges (much brighter than the stellar sample considered as contaminants) do not make them good candidates for deriving this uncertainty.}
    \label{fig:starmap_comparison}
\end{figure}

One of the components of these calculations is the choice of a star map to establish the density relationships. We have derived a $\sim1\%$ systematic uncertainty in the estimation of the impurity derived from comparing brighter and fainter stellar samples (Figure \ref{fig:starmap_comparison}). The 2MASS and Tycho-2 \citep{2mass,tycho2} stellar maps are included for completeness, but their magnitude range does not track accurately the range of brightness we need to account for Milky Way distribution in DES. Gaia's DR2 corresponds to the data described in \citet{gaiadr2}.

\subsection{Galaxy ratio near the Large Magellanic Cloud}

\begin{figure}
	\includegraphics[width=\columnwidth]{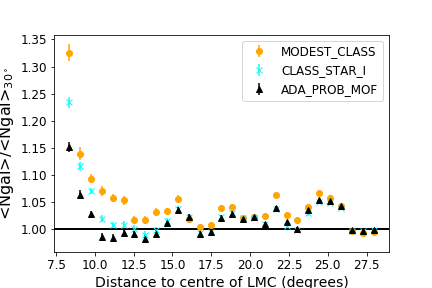}
    \caption{Galaxy ratio (with respect to galaxy density at 30 degrees from LMC) for as a function of angular distance from the LMC centre.}
    \label{fig:lmc_test}
\end{figure}

Using the same pixelisation as above, we also approach the comparison of different classifiers using a figure of merit based on the identified galaxy density in each of these pixels, as compared to the one found at a certain distance to the centre of the Large Magellanic Cloud (LMC), set at ($\alpha$,$\delta$) = (5h23m34.5s, $-69^\circ$45'11"). This value is normalised to one at 30 degrees from the centre of the LMC (Figure \ref{fig:lmc_test}). Here we use a flux limited sample with $i<23$. In this case, we can see a clear advantage in using a classifier with multiple input attributes (including colour), possibly helped by the fact that in a crowded field such as the peripheries of the LMC, morphology starts to have a smaller discriminating power. On the other hand, the LMC has a bluer population, but this doesn't seem to offset the ML classification significantly, though this aspect is worth studying further in a future work.

Using a metric such as this at a given fixed distance of the LMC could be useful as a figure of merit. In this case 10 degrees seems convenient but we must remark that this could be due to the odd geometry available around the LMC, so other photometric surveys might find other ranges for comparison more valuable.

\subsection{Stellar locus of classified stars}

\begin{figure}
	\includegraphics[width=\columnwidth]{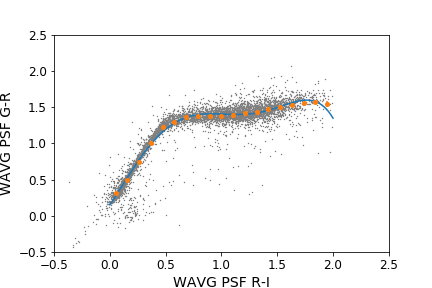}
    \caption{Fit to stellar locus using a fifth degree polynomial}
    \label{fig:slocus_cosmos}
\end{figure}

\begin{figure}
	\includegraphics[width=\columnwidth]{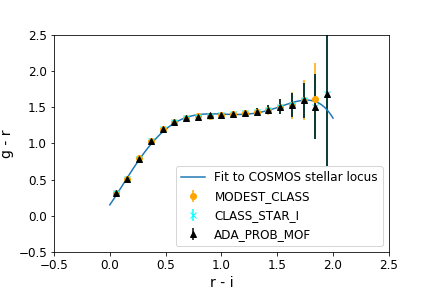}
	\includegraphics[width=\columnwidth]{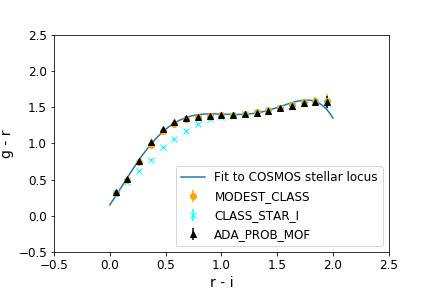}
    \caption{Stellar locus for star samples from various classifiers, \NEW{for a bright sample  ($i<21$, \textit{top}) and a fainter one ($i<24$, \textit{bottom})}.}
    \label{fig:slocus}
\end{figure}

Finally, we tested the consistency of the stellar locus derived in $r-i$ vs. $g-r$ colour space to a similar fit to stars in the COSMOS field. The stellar locus was fit by a fifth-order polynomial, as shown in Figure \ref{fig:slocus_cosmos}, similarly to what is realised in \citet{covey}. 
The same fit curve from Figure \ref{fig:slocus_cosmos} is shown again versus several classifiers in Figure \ref{fig:slocus}. In general a good agreement is seen except for the faintest end, where classified stars seem to deviate from the expected stellar locus for {\tt CLASS\_STAR}. 

 

\section{Discussion: implications for large-scale structure and Milky Way studies}
\label{sec:discussion}

In the previous section we explored a variety of tests both with and without truth information assessing the relative performance of a wide range of star-galaxy classifiers in DES Y1 data. We now turn to the impact of making different selections on scientific analyses of interest to astronomers and cosmologists. Though it is beyond the scope of this work to define specific choices for any arbitrary  study, in this section we sketch out the general implications of the results shown here for two broad ranging topics of interest, namely the large-scale structure (LSS) of galaxies and Milky Way analyses within DES. With regards to weak lensing shear catalogues, \citet{zuntz} have shown that star-galaxy contamination is at most a second-order contaminant when either {\tt MODEST\_CLASS} or {\tt MCAL\_RATIO} are used
for the DES Y1 cosmology analyses.
For a thorough discussion on LSS and weak lensing requirements for star-galaxy separation, \citet{soumagnac} provides an in-depth review. 

\subsection{Large-Scale Structure}
\label{sec:lss}

The impact of stellar contamination on studies of clustering amplitude has been well studied for several years now \citep[e.g.,]{ross,crocce} with an impact of the order of $(1-I)^2$ in the angular correlation function $\omega(\theta)$ if we assume an unclustered component that contaminates the galaxy population with impurity fraction $I$. A large contamination can severely dilute the signal (reducing the significance of the BAO peak as shown by \citet{carnero}), or even create a large-scale component if unaccounted for, thus mimicking an effect such as primordial non-Gaussianities \citep{giannantonio}. However, in the range $I\sim$ O($2\%$), the accuracy by which we determine $I$ becomes much more relevant, as this is the systematic that will dominate in the determination of the uncertainty in galaxy bias measurements and multiple probe analyses. 

\begin{figure}
	\includegraphics[width=\columnwidth]{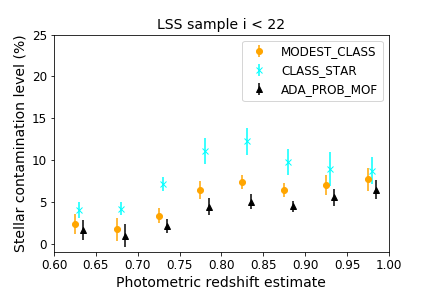}
	\includegraphics[width=\columnwidth]{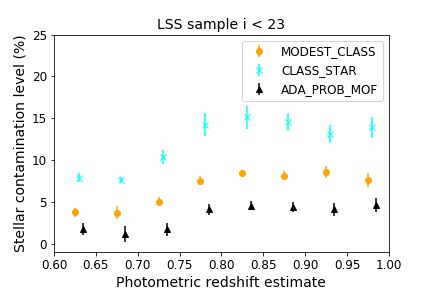}
    \caption{Stellar contamination level as a function of redshift \NEW{for a bright sample  (\textit{top}, i$<22$) and a faint sample (\textit{bottom}, i$<23$), derived with the method described in Section \ref{sec:densvsdens} for different samples classified by photometric redshift.}}
    \label{fig:lss_sample_test}
\end{figure}

Figure \ref{fig:classcomp_densdens} implies that the choice of classifier does not matter too much for cosmology analyses in the broadest sense. However, going into a more realistic sample for large scale structure studies, using a selection for red galaxies that have better estimated photo-z and galaxy bias \citep{y1baosample} for BAO analysis for example, some evident differences appear for the highest redshifts (where due to their colours, many faint stars are mis-classified into those bins of photo-z). This is the main photo-z region of interest for BAO for DES. Also between the classifiers, which become more evident when the flux cut is driven to fainter magnitudes as shown before. See Figure \ref{fig:lss_sample_test}.  

\begin{figure}
	\includegraphics[width=\columnwidth]{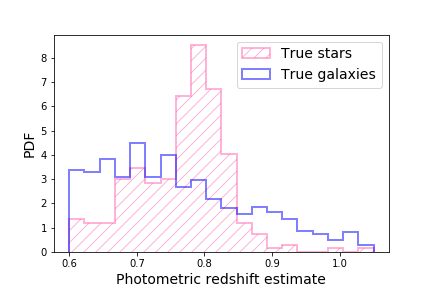}
    \caption{Normalised distribution of BPZ redshifts for a typical red galaxy sample that would be used for LSS studies, over a region with known identification of stars and galaxies through Hubble Space Telescope imaging.}
    \label{fig:bpz_hsc}
\end{figure}

These results show that a realistic LSS sample, is more severely affected by stellar contamination, driving the impurity levels up to $5-6\%$ in some redshift bins. This is seen more clearly in Figure \ref{fig:bpz_hsc} where photo-zs are shown for the true stars in the fields overlapping the COSMOS region for a general selection and an LSS-like, red galaxy selection. One way to drive down this impurity therefore is to either apply more stringent constraints to the star-galaxy thresholds, sacrificing a percentage of true galaxies along the way. For the case of {\tt MODEST\_CLASS} and {\tt ADA\_PROB\_MOF}, we can push down to 2\% by removing ${\sim}9\%$ and ${\sim}4\%$ galaxies respectively. Though a ML approach seems more convenient in this case, the use of colour and magnitude information may lead to potential correlations between object classification and photo-z determination that must be investigated in more detail. As for the uncertainty of determining $I$ using the density plots, Figure \ref{fig:starmap_comparison} shows that using fainter stellar maps to derive the impurity via this method generates a different contamination rate. This can be due to tracing of different components of the Galaxy, but for maps built upon possibly contaminated data it could well be that the star maps themselves are not ideal (e.g. the bright {\tt MODEST\_CLASS} stars could have a small component from misclassified compact galaxies). An improvement in understanding the underlying Galactic stellar structure through simulations or an adequate culling of the reference stellar maps to improve agreement would reduce this limitation in the determination of the impurity level, $I$.

\subsection{Milky Way}
\label{sec:mw}

\begin{figure}
	\includegraphics[width=\columnwidth]{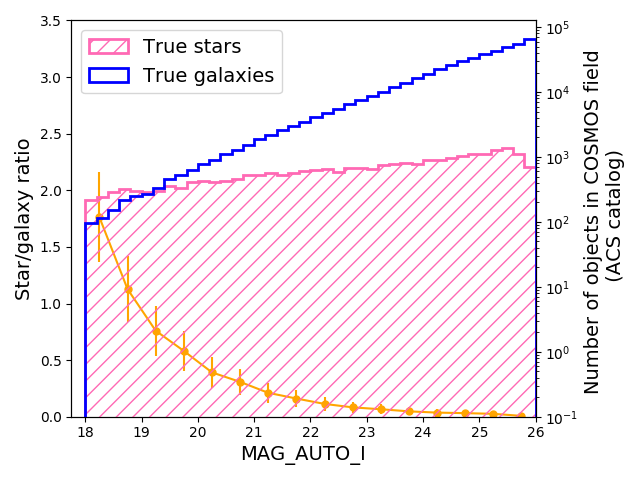}
    \caption{Star-galaxy ratio in differential {\tt MAG\_AUTO} bins, taken from the COSMOS ACS catalogue. Point-sources are overwhelmed by extended sources in the faint end.}
    \label{fig:sgratio}
\end{figure}

In the case of Milky Way studies, in broad terms we are interested in obtaining a more complete and pure stellar sample, down to faint magnitudes. Studies such as those in \citet{fadely}, show that currently this can become a major systematic effect in deriving the Galaxy structure. Additionally, misclassified galaxies become a limiting factor for discovering faint resolved stellar overdensities \citep[e.g.,]{willman,bechtol,y2satellites,pieres}. This problem is evidenced returning to the COSMOS ACS catalogue used in Section \ref{sec:calib}, which can be used to understand the ratio of stars to galaxies up to a very faint limit (shown in Figure \ref{fig:sgratio}).

In this sense, the results in \citet{pieres} or \citet{shipp}, for example, show that the very good results can be obtained based on a multi-epoch based classifier such as the weighted averaged {\tt SPREAD\_MODEL} quantity or the {\tt MOF} pipeline. 

The use of machine learning codes in this case is limited by the fact that if we want to study the distribution of specific types of stars, or search for Milky Way neighbours with a particular range of colours and magnitudes, we have to be very careful with introducing biases or complex selection functions in our application sample, much like what happens with photometric redshifts for the LSS case.

What the results of the current study show (e.g. Figure \ref{fig:pr_snfields_stars}) is that the {\tt MOF} technique has the potential of being the best candidate for selecting stellar candidates from its very tight morphological stellar locus and its capacity of reaching deeper into the separation of extended and point-like sources, by increasing by $\sim20\%$ the amount of stars in the sample for a given purity and magnitude cut versus a `classical' {\tt SPREAD\_MODEL} cut (in this plot, at 0.8 purity we go from 0.70 to 0.84 completeness). However, additional fine-tuning of the algorithm is needed to reach a good completeness in the bright end, where the model fit is not especially attuned to fits of stellar shapes. This is an open line of development in the algorithm in DES.

\section{Conclusions}
\label{sec:conclusions}
In this paper, we have compiled a wide variety of tests over a diverse array of star-galaxy classifiers for the DES Y1 dataset. These tests can be ported or used as examples for any other photometric dataset. The classifiers range from well-tested algorithms in the literature, to new developments using morphological information and/or flux information, using priors for stars/galaxies or training sets for machine learning codes based on space imaging information from the Hubble Space Telescope. We have studied their relative performance both using accurate truth information from spectroscopic and space imaging external datasets, and devised tests over the broad DES Y1 footprint that do not require this information. In the light of these results, we have analyzed the impact of using these algorithms on two broad science cases of interest to users of the DES data, namely, large-scale structure analyses and Milky Way studies. Star-galaxy classification remains as a non-dominant but important systematic source of error for cosmology, and very critical for Milky Way structure measurements and discoveries. These are the specific items that were highlighted in this work:

\begin{itemize}
\item Machine learning methods perform very well on calibration fields tests (Figures \ref{fig:roc_snfields} to \ref{fig:pr_snfields_stars} and Table \ref{tab:aucs}). In the application field the results are slightly better than for non-ML classification, especially in the faint end (Figure \ref{fig:lss_sample_test}). Optical colour based classifiers however could potentially introduce biases in sample selection.
\item Although {\tt CLASS\_STAR} has been used in the past to good effect, its lack of performance in the faint end (see e.g. Figures \ref{fig:roc_cosmos} and \ref{fig:nm_stars}) leads us to recommend alternative classification methods such as {\tt SExtractor}'s {\tt SPREAD\_MODEL} or a multi-epoch fit to the shape. In this sense, using multi-epoch, multi-object fitting instead of directly using coadded information is the preferred option for object classification in optical wavelengths (as shown in Section \ref{sec:calib}).
\item As has been demonstrated in the past, the addition of infrared data is very valuable, albeit limited currently by the depth and extension of such surveys (Section \ref{sec:spectralcalib}).
\item Photometric redshift binning will affect stellar contamination of specific galaxy samples (Figure \ref{fig:bpz_hsc}).
\end{itemize}

\subsection{Expected improvements for Y3 and beyond}

Considering these results, we have identified very clear future directions to expand and improve star-galaxy classification in forthcoming DES science analyses (Y3 and beyond).

\begin{itemize}
\item Improvement of the \texttt{MOF} quantities to better fit stellar shapes and prevention of fitting failures. 
\item Understanding the impact of using colour information on specific science cases (photo-z, stellar type selections) to ascertain whether or not the usage of this information in ML codes hampers their utility for star-galaxy separation in extragalactic and Milky Way studies respectively, in exchange of an additional 2-5\% in purity depending on the case.
\item The combination of information as done in \citet{kim} from different approaches, especially adding external infrared colours, could greatly benefit the performance of some classifiers. Once an adequate template set is studied for the DES data, trying to overcome the impact of the lack of u-band information, template-based codes could be considered as well to complement this impact study. In addition, this would provide a truly probabilistic output that could be employed in statistical studies of large-scale structure, removing the need of having to eliminate a subsample of galaxies according to an arbitrary threshold.
\item Besides VHS data, the addition of Gaia's DR2 information \citep{gaiadr2} will provide a robust and broad complement to these tests at magnitudes $r<21$.
\end{itemize}

\subsection{Ideas for further study}

Finally, we call attention to other approaches and tests that we have not specifically investigated here which could be relevant for future studies:

\begin{itemize}
\item Adding available $u$-band and \NEW{specially infrared band information using matched-aperture photometry as part of the algorithms used here}.
\item With respect to a template-fitting approach, the characteristics of this dataset (lack of $u$-band or infrared information), severely limit its usability. But expanding the dataset, jointly with an accurate understanding of the template range to be used can be considered as a promising approach if these requirements are met, to be used in a joint probabilistic method.
\item Including very detailed image-based simulations for training, such as {\tt Balrog} \citep{balrog} or {\tt UFIG} \citep{ufig}, to understand the failure modes of different classifiers.
\item Adding seeing as part of the features of the machine learning classifiers, as well as for characterization of the performance of the different approaches.
\item Usage of the object position in the sky can also provide an additional lever for a probabilistic approach, as a prior to be added to the overall posterior estimation. This should be approached with care for certain analysis (e.g. Milky Way structure).
\item PSF homogeneization will improve the \texttt{SExtractor} estimates as shown in \citet{desai}. However, using {\tt MOF}-based photometry is a more promising alternative that avoids some of the problems associated with homogeneization.
\item Convolutional Neural Networks \citep[e.g.,][]{kimdcnn} can be applied directly to the images to provide a new and complementary approach to ML applied at catalogue-level. Image-level analyses may benefit by using information from multiple (>10) bands (e.g., \citet{cabayol}). 
\end{itemize}

The data used in this paper are provided at \NEW{\url{http://des.ncsa.illinois.edu/releases/y1a1}}.

\section*{Acknowledgements}

ISN would like to thank \v{Z}.Ivezi\'c and A.Robin for useful discussions and insights; R.Gonz\'alez-G\'erboles for help in carrying out the HB tests; S.Koposov for providing useful insights into the expected stellar distributions; F.Ostrovsky for expert opinion on star-QSO classification and possible impact and A.Kov\`acs for suggestions on using infrared datasets. 

E.Balbinot acknowledges financial support from the European Research Council (StG-335936).

Funding for the DES Projects has been provided by the U.S. Department of Energy, the U.S. National Science Foundation, the Ministry of Science and Education of Spain, 
the Science and Technology Facilities Council of the United Kingdom, the Higher Education Funding Council for England, the National Center for Supercomputing 
Applications at the University of Illinois at Urbana-Champaign, the Kavli Institute of Cosmological Physics at the University of Chicago, the Center for Cosmology and Astro-Particle Physics at the Ohio State University,the Mitchell Institute for Fundamental Physics and Astronomy at Texas A\&M University, Financiadora de Estudos e Projetos, Funda{\c c}{\~a}o Carlos Chagas Filho de Amparo {\`a} Pesquisa do Estado do Rio de Janeiro, Conselho Nacional de Desenvolvimento Cient{\'i}fico e Tecnol{\'o}gico and 
the Minist{\'e}rio da Ci{\^e}ncia, Tecnologia e Inova{\c c}{\~a}o, the Deutsche Forschungsgemeinschaft and the Collaborating Institutions in the Dark Energy Survey. 

The Collaborating Institutions are Argonne National Laboratory, the University of California at Santa Cruz, the University of Cambridge, Centro de Investigaciones Energ{\'e}ticas, 
Medioambientales y Tecnol{\'o}gicas-Madrid, the University of Chicago, University College London, the DES-Brazil Consortium, the University of Edinburgh, 
the Eidgen{\"o}ssische Technische Hochschule (ETH) Z{\"u}rich, 
Fermi National Accelerator Laboratory, the University of Illinois at Urbana-Champaign, the Institut de Ci{\`e}ncies de l'Espai (IEEC/CSIC), the Institut de F{\'i}sica d'Altes Energies, Lawrence Berkeley National Laboratory, the Ludwig-Maximilians Universit{\"a}t M{\"u}nchen and the associated Excellence Cluster Universe, the University of Michigan, the National Optical Astronomy Observatory, the University of Nottingham, The Ohio State University, the University of Pennsylvania, the University of Portsmouth, SLAC National Accelerator Laboratory, Stanford University, the University of Sussex, Texas A\&M University, and the OzDES Membership Consortium.

Based in part on observations at Cerro Tololo Inter-American Observatory, National Optical Astronomy Observatory, which is operated by the Association of 
Universities for Research in Astronomy (AURA) under a cooperative agreement with the National Science Foundation.

The DES data management system is supported by the National Science Foundation under Grant Numbers AST-1138766 and AST-1536171. The DES participants from Spanish institutions are partially supported by MINECO under grants AYA2015-71825, ESP2015-66861, FPA2015-68048, SEV-2016-0588, SEV-2016-0597, and MDM-2015-0509, some of which include ERDF funds from the European Union. IFAE is partially funded by the CERCA program of the Generalitat de Catalunya.Research leading to these results has received funding from the European Research Council under the European Union's Seventh Framework Program (FP7/2007-2013) including ERC grant agreements 240672, 291329, and 306478.
We  acknowledge support from the Australian Research Council Centre of Excellence for All-sky Astrophysics (CAASTRO), through project number CE110001020, and the Brazilian Instituto Nacional de Ci\^encia e Tecnologia (INCT) e-Universe (CNPq grant 465376/2014-2).

This manuscript has been authored by Fermi Research Alliance, LLC under Contract No. DE-AC02-07CH11359 with the U.S. Department of Energy, Office of Science, Office of High Energy Physics. The United States Government retains and the publisher, by accepting the article for publication, acknowledges that the United States Government retains a non-exclusive, paid-up, irrevocable, world-wide licence to publish or reproduce the published form of this manuscript, or allow others to do so, for United States Government purposes.

This research uses data from SDSS-III. Funding for SDSS-III has been provided by the Alfred P. Sloan Foundation, the Participating Institutions, the National Science Foundation, and the U.S. Department of Energy Office of Science. The SDSS-III web site is \url{http://www.sdss3.org/}. SDSS-III is managed by the Astrophysical Research Consortium for the Participating Institutions of the SDSS-III Collaboration including the University of Arizona, the Brazilian Participation Group, Brookhaven National Laboratory, Carnegie Mellon University, University of Florida, the French Participation Group, the German Participation Group, Harvard University, the Instituto de Astrofisica de Canarias, the Michigan State/Notre Dame/JINA Participation Group, Johns Hopkins University, Lawrence Berkeley National Laboratory, Max Planck Institute for Astrophysics, Max Planck Institute for Extraterrestrial Physics, New Mexico State University, New York University, Ohio State University, Pennsylvania State University, University of Portsmouth, Princeton University, the Spanish Participation Group, University of Tokyo, University of Utah, Vanderbilt University, University of Virginia, University of Washington, and Yale University.

This research uses data from the VIMOS VLT Deep Survey, obtained from the VVDS database operated by Cesam, Laboratoire d'Astrophysique de Marseille, France.

This research uses data based on observations made with the NASA/ESA Hubble Space Telescope, and obtained from the Hubble Legacy Archive, which is a collaboration between the Space Telescope Science Institute (STScI/NASA), the Space Telescope European Coordinating Facility (ST-ECF/ESAC/ESA) and the Canadian Astronomy Data Centre (CADC/NRC/CSA). 

This research uses data based on zCOSMOS observations carried out	using the Very Large Telescope at the ESO Paranal Observatory under	Programme ID: LP175.A-0839.

The VISTA Data Flow System pipeline processing and science archive are described in \citet{irwin}, \citet{hambly} and \citet{cross}.

This work has made use of data from the European Space Agency (ESA) mission {\it Gaia} (\url{https://www.cosmos.esa.int/gaia}), processed by the {\it Gaia} Data Processing and Analysis Consortium (DPAC,\url{https://www.cosmos.esa.int/web/gaia/dpac/consortium}). Funding for the DPAC
has been provided by national institutions, in particular the institutions participating in the {\it Gaia} Multilateral Agreement.

CosmoHub has been developed by the Port d'Informaci\'o Cient\'ifica (PIC), maintained through a collaboration of the Institut de F\'isica d'Altes Energies (IFAE) and the Centro de Investigaciones Energ\'eticas, Medioambientales y Tecnol\'ogicas (CIEMAT). The work was partially funded by the "Plan Estatal de Investigaci\'on Cient\'ifica y T\'ecnica y de Innovaci\'on" program of the Spanish government.




\bibliographystyle{mnras}
\bibliography{sgy1} 




\appendix

\section{{\tt ADA\_PROB} technical details}
\label{sec:ada_prob_desc}

This appendix describes the details of one of the machine learning frameworks called {\tt ADA\_PROB}.

The framework first selects an exhaustive list of photometric properties, or features, and generates linear combinations of these features to produced new features. This may include unphysical combinations, such as magnitudes and radii being combined. We also generate features `intelligently', by using the current state of the art. For the problem of star-galaxy separation for DES, this means including both a binary {\tt MODEST\_CLASS} class value, and a continuous {\tt MODEST\_CLASS} variable for both stars and galaxies.

Next, the enormous feature list is sorted by rank, using the value of the mutual information\footnote{\url{https://en.wikipedia.org/wiki/Mutual_information}}, which is a non-linear correlation coefficient, between the selected feature and the target class. Finally the top 150 features are selected to form the inputs to the machine learning algorithms. 

The framework then explores many machine learning algorithms, each of which are trained with random variations of each of their own hyper-parameters. The framework explores a plethora of algorithms, drawn from the sci-kit-learn \citep{scikit-learn} package. These include AdaBoost, which often performs well, and also Random Forests, Extra Randomised Trees, Quadratic Discriminant Analysis and the K-Nearest Neighbours Classifier. 

The performance of each selected algorithm and set of hyper-parameters is quantified by measuring the average $F_1$ score on 30 held out samples during 30 fold cross validation. The $F_1$ score is the geometric mean between the precision and the recall, and 30 fold cross validation is akin to making 30 jackknife samples of the data, training on all but the held out sample, and then making predictions on that held out sample, and then repeating. The held out jackknife results, or ``class weights", for each training object are retained for future classification calibration. 

The winning algorithm and hyper-parameter set is then retrained on the full training sample. The training procedure is deemed to have been completed once at least 50 systems have been explored and when the $F_1$ score has not been improved upon after 20 iterations. In our empirical experience, we find this to be a generally stable point at which one can stop the exploration of the different algorithms, hyper-parameters, and move on to the final stage of the framework. 

This final stage then uses  isotonic regression to calibrate the held out class weights of the training data. This enforces the statistical properties of the class weights to more closely resemble a probability. This rescaling is performed by comparing the total number of those objects within a class weight bin, with the fraction of objects to have the true class value. This comparison leads to a rescaling of class weights to class probabilities which we note are conditional on the training data.

The winning machine learning algorithm, which happened to be AdaBoost in this case, is then used to make class weight predictions on both the test sample and the science samples, and their output class weights are scaled using the previously learned rescaling, to make them more closely resemble probabilities.

We can also perform a feature importance analysis \citep[see, e.g.,][]{2015MNRAS.449.1275H} which suggests that the features with the most predictive power are indeed those derived from {\tt MODEST}, with other ranking features being  {\tt WAVG\_SPREAD\_MODEL\_R} and {\tt MAGERR\_MODEL\_I}.

\section{External datasets}
\label{sec:external_datasets}
\subsection{Access to external catalogues used in this work}
The catalogues used in Section \ref{sec:calib} and listed on Table \ref{tab:external_datasets} can be obtained from the following website: \url{http://des.ncsa.illinois.edu/releases/y1a1}.
\subsection{Queries used to extract the datasets}
Query to the SDSS CASJOBS interface (used as imaging truth table for some tests) to obtain 2MASS, WISE matches with SDSS data, to match with DES data on same area.
\begin{verbatim}
SELECT
  s.ra, s.dec, s.dered_r, 
  w.w1mpro as w1, w.j_m_2mass as j, s.z, 
  s.class
INTO 
  mydb.stripe82_wise_2mass_z_match
FROM
  wise_xmatch as xm
JOIN 
   specPhoto as s on xm.sdss_objid = s.objid
JOIN 
    wise_allsky as w on xm.wise_cntr = w.cntr
WHERE
   ((s.dered_g < 23.0) or (s.dered_r < 23.0) 
   or (s.dered_i < 23.0)) and
   ((s.ra > 0 and s.ra < 5 and s.dec > -2.5 and 
   s.dec < 3.5) or  (s.ra > 315 and s.dec > -3 
   and s.dec < 3)) and s.zWarning = 0 
   and s.zErr < 0.001
      
\end{verbatim}

\noindent \NEW{Query to the SDSS CASJOBS interface (used as imaging truth table for some tests) to match with DES and VHS data on same area.}
\begin{verbatim}
SELECT
  s.ra, s.dec, s.dered_r, s.z, s.class
INTO 
  mydb.stripe82_z_dr13
FROM
  specPhoto as s
WHERE
  ((s.ra > 0 and s.ra < 5 and 
  s.dec > -2.5 and s.dec < 3.5) or
    (s.ra > 315 and s.dec > -3 and 
    s.dec < 3)) and s.zWarning = 0
  and s.zErr < 0.001 
\end{verbatim}

\noindent Query to the Hubble Source Catalog CASJOBS interface (used as imaging truth table for some tests):
\begin{verbatim}
SELECT
     p.MatchRA, p.MatchDEC, p.MatchID as hscv2_id, 
     p.CI, p.CI_Sigma, m.A_F814W, m.A_F814W_Sigma
INTO
     hsc_source_catalog
FROM 
     SumPropMagAutoCat p 
JOIN
     SumMagAutoCat m ON p.MatchID = m.MatchID
WHERE
     m.A_F814W > 0 and m.A_F814W_Sigma is not null 
     and p.numimages > 2
\end{verbatim}

\noindent Query to the VISTA Science Archive, using the VHSDR3 database.
\begin{verbatim}
SELECT     ra, dec, jpetromag, jpetromagerr, 
jmksext, jmksexterr
FROM 
     vhsSource
WHERE 
     jerrbits = 0 and kserrbits = 0 and 
     (priOrSec=0 OR priOrSec=frameSetID) and
     dec between -2 and 2 and (ra > 315 or ra < 5) 
\end{verbatim}

\noindent Query to Gaia's DR2, using the CosmoHub \citep{cosmohub} interface.
\begin{verbatim}
SELECT `ra`, `dec`, `phot_g_mean_mag`, `l`, `b`, 
`phot_g_mean_flux_over_error`,
`astrometric_primary_flag` 
FROM gaia_dr2 
WHERE
((`ra` > 305) or (`ra` < 90)) and (`dec` > -61) 
and (`dec` < -35) and phot_g_mean_mag > 18.5 
\end{verbatim}


\section*{AFFILIATIONS}
\label{sec:affiliations}
$^{1}$ Centro de Investigaciones Energ\'eticas, Medioambientales y Tecnol\'ogicas (CIEMAT), Madrid, Spain\\
$^{2}$ Max Planck Institute for Extraterrestrial Physics, Giessenbachstrasse, 85748 Garching, Germany\\
$^{3}$ Universit\"ats-Sternwarte, Fakult\"at f\"ur Physik, Ludwig-Maximilians Universit\"at M\"unchen, Scheinerstr. 1, 81679 M\"unchen, Germany\\
$^{4}$ Department of Physics \& Astronomy, University College London, Gower Street, London, WC1E 6BT, UK\\
$^{5}$ Department of Particle Physics and Astrophysics, Weizmann Institute of Science, Rehovot 76100, Israel\\
$^{6}$ LSST, 933 North Cherry Avenue, Tucson, AZ 85721, USA\\
$^{7}$ Fermi National Accelerator Laboratory, P. O. Box 500, Batavia, IL 60510, USA\\
$^{8}$ Department of Physics and Electronics, Rhodes University, PO Box 94, Grahamstown, 6140, South Africa\\
$^{9}$ Institut de F\'{\i}sica d'Altes Energies (IFAE), The Barcelona Institute of Science and Technology, Campus UAB, 08193 Bellaterra (Barcelona), Spain \\
$^{10}$Kavli Institute for Cosmological Physics, University of Chicago, Chicago, IL 60637, USA\\
$^{11}$ Department of Physics, University of Surrey, Guildford GU2 7XH, UK\\
$^{12}$ Institute of Astronomy, University of Cambridge, Madingley Road, Cambridge CB3 0HA, UK\\
$^{13}$ Kavli Institute for Cosmology, University of Cambridge, Madingley Road, Cambridge CB3 0HA, UK\\
$^{14}$ CNRS, UMR 7095, Institut d'Astrophysique de Paris, F-75014, Paris, France\\
$^{15}$ Sorbonne Universit\'es, UPMC Univ Paris 06, UMR 7095, Institut d'Astrophysique de Paris, F-75014, Paris, France\\
$^{16}$ Department of Astronomy, University of Illinois at Urbana-Champaign, 1002 W. Green Street, Urbana, IL 61801, USA\\
$^{17}$ National Center for Supercomputing Applications, 1205 West Clark St., Urbana, IL 61801, USA\\
$^{18}$ Center for Cosmology and Astro-Particle Physics, The Ohio State University, Columbus, OH 43210, USA\\
$^{19}$ Kavli Institute for Particle Astrophysics \& Cosmology, P. O. Box 2450, Stanford University, Stanford, CA 94305, USA\\
$^{20}$ SLAC National Accelerator Laboratory, Menlo Park, CA 94025, USA\\
$^{21}$Instituto de F\'\i sica, UFRGS, Caixa Postal 15051, Porto Alegre, RS - 91501-970, Brazil\\
$^{22}$Laborat\'orio Interinstitucional de e-Astronomia - LIneA, Rua Gal. Jos\'e Cristino 77, Rio de Janeiro, RJ - 20921-400, Brazil\\
$^{23}$Brookhaven National Laboratory, Bldg 510, Upton, NY 11973, USA\\
$^{24}$Department of Physics, University of Chicago, Chicago, Illinois 60637, USA
$^{25}$Cerro Tololo Inter-American Observatory, National Optical Astronomy Observatory, Casilla 603, La Serena, Chile\\
$^{26}$Observat\'orio Nacional, Rua Gal. Jos\'e Cristino 77, Rio de Janeiro, RJ - 20921-400, Brazil\\
$^{27}$Department of Physics, IIT Hyderabad, Kandi, Telangana 502285, India\\
$^{28}$Instituto de Fisica Teorica UAM/CSIC, Universidad Autonoma de Madrid, 28049 Madrid, Spain\\
$^{29}$Institut d'Estudis Espacials de Catalunya (IEEC), 08193 Barcelona, Spain\\
$^{30}$Institute of Space Sciences (ICE, CSIC),  Campus UAB, Carrer de Can Magrans, s/n,  08193 Barcelona, Spain\\
$^{31}$Santa Cruz Institute for Particle Physics, Santa Cruz, CA 95064, USA\\
$^{32}$Department of Physics, The Ohio State University, Columbus, OH 43210, USA\\
$^{33}$Harvard-Smithsonian Center for Astrophysics, Cambridge, MA 02138, USA\\
$^{34}$Department of Astronomy/Steward Observatory, 933 North Cherry Avenue, Tucson, AZ 85721-0065, USA\\
$^{35}$Jet Propulsion Laboratory, California Institute of Technology, 4800 Oak Grove Dr., Pasadena, CA 91109, USA\\
$^{36}$Australian Astronomical Observatory, North Ryde, NSW 2113, Australia\\
$^{37}$Departamento de F\'isica Matem\'atica, Instituto de F\'isica, Universidade de S\~ao Paulo, CP 66318, S\~ao Paulo, SP, 05314-970, Brazil\\
$^{38}$Department of Physics and Astronomy, University of Pennsylvania, Philadelphia, PA 19104, USA\\
$^{39}$Instituci\'o Catalana de Recerca i Estudis Avan\c{c}ats, E-08010 Barcelona, Spain\\
$^{40}$Department of Physics, University of Michigan, Ann Arbor, MI 48109, USA\\
$^{41}$School of Physics and Astronomy, University of Southampton,  Southampton, SO17 1BJ, UK\\
$^{42}$Brandeis University, Physics Department, 415 South Street, Waltham MA 02453, USA\\
$^{43}$Instituto de F\'isica Gleb Wataghin, Universidade Estadual de Campinas, 13083-859, Campinas, SP, Brazil\\
$^{44}$Computer Science and Mathematics Division, Oak Ridge National Laboratory, Oak Ridge, TN 37831, USA \\
$^{45}$Institute of Cosmology \& Gravitation, University of Portsmouth, Portsmouth, PO1 3FX, UK\\

\bsp	
\label{lastpage}
\end{document}